\documentclass[aip, reprint]{revtex4-2}
\usepackage[T1]{fontenc}
\usepackage[russian,english]{babel}
\usepackage{graphicx}
\usepackage{amsmath}
\usepackage{amssymb}
\usepackage{mathtools}
\usepackage{physics}
\usepackage{color,soul}
\usepackage[colorlinks=true,allcolors=blue]{hyperref}
\usepackage{cancel}
\usepackage{xfrac}
\usepackage{comment}
\usepackage{siunitx}

\usepackage{upgreek}
\usepackage[dvipsnames]{xcolor}
\usepackage{microtype}
\usepackage{xcolor}

\graphicspath{{./Figs/}}
\begin{document}
\title{High-order corrections to the radiation-free dynamics of an electron in the strongly radiation-dominated regime}

\date{\today}
\author{A.~S.~Samsonov}
\email[Corresponding author: ]{asams@ipfran.ru}
\author{E.~N.~Nerush}
\author{I.~Yu.~Kostyukov}
\affiliation{Institute of Applied Physics of the Russian Academy of
    Sciences, 46 Ulyanov St., Nizhny Novgorod 603950, Russia}

\begin{abstract}
    A system of reduced equations is proposed for the electron motion in the strongly-radiation dominated regime for an arbitrary electromagnetic field configuration.
    The developed approach is used to analyze various scenarios of an electron dynamics in the strongly-radiation dominated regime: motion in rotating electric and magnetic fields, longitudinal acceleration in a plane wave and in a plasma wakefield.
    The obtained results show that the developed approach is able to describe features of the electron dynamics, which are essential to a certain scenario, but which could not be captured in the framework of the original radiation-free approximation [A.~S.~ Samsonov et al., Phys.~Rev.~A \textbf{98}, 053858 (2018); A.~Gonoskov and M.~Marklund, Phys. Plasmas \textbf{25}, 093109 (2018)].
    The results are verified by numerical integration  of non-reduced motion equations with account of radiation reaction in both semi-classical and fully quantum cases.
\end{abstract}

\maketitle

\section{Introduction}
\label{sec.Intro}
With upcoming laser facilities, such as ELI~\cite{ELI}, SULF~\cite{SULF}, SEL~\cite{SEL}, XCELS~\cite{XCELS}, etc.~investigation of laser-matter interaction in the regime of extreme intensity will become feasible.
Radiation reaction is expected to accompany such interaction, although, its direct impact is usually quite hard to predict.
The fact that the charged particle experiences a recoil force when radiating has been known for more than a century, however a consistent model describing this phenomenon in both theoretical and numerical studies is still being debated.
Recently conducted experiments aimed at determining the correct model of radiation friction still contain a certain level of ambiguity~\cite{cole2018, poder2018}, and thus cannot alleviate the problem.
The problem becomes more and more acute with growing number of studies discovering possible new effects caused by radiation reaction.
Apparently, these effects vary greatly and include e.g.~alteration of the particle acceleration mechanisms~\cite{Tamburini10, Tamburini12, kostyukov2012radiative, Capdessus12, Capdessus15, Nerush15, Gelfer18a, Gelfer18b, gelfer2021ions, golovanov2021radiation}, highly efficient laser pulse absorption~\cite{Grismayer16}, relativistic transparency reduction~\cite{Zhang15}, inverse Faraday effect~\cite{Liseykina16,liseykina2021IFE,Samsonov2021IFE}, particles polarization~\cite{DelSorbo2017Spin,DelSorbo2018spin,Chen2019spin,Seipt2019spin,Wu2019spin,Li2019spin,Li2020spin,Wan2020spin,gong2021spin}, initiating quantum electrodynamical (QED) cascades~\cite{nerush2007radiation,Bell2008,Nerush2011,Ridgers2012,narozhny2015quantum,Kostyukov2016,grismayer2017seeded,jirka2017qed,luo2018qed,Yuan2018,del2018ion,Lu2018,Luo2018,efimenko2019laser,samsonov2019laser, samsonov2021hydrodynamical} and many others.
Signatures of these effects are expected to be most prominent in the so called radiation-dominated regime, i.e.~regime when radiative losses of charged particles are comparable with energy gain in the EM field.
Estimates show that field amplitudes needed for realization of this regime can be achieved experimentally at either future laser facilities such as ELI, SEL, XCELS, or future accelerators such as FACET-II~\cite{FACET}.

Via QED one can describe radiation reaction self-consistently and calculate probability of radiating a photon with a given energy.
While full QED description is the most accurate description of the radiation reaction at the moment, it is not usually applicable to practical problems involving complex light-matter interactions, since one calculates scattering probabilities between some stationary (commonly Volkov) electron states.
To describe a dynamic problem where these states evolve due to evolution of EM fields, non-stationary Dirac equations have to be solved, which is usually either unfeasible or impractical.
However, in some conditions this is not necessary, since the problem can be significantly simplified.
The first main parameter which define such condition is the dimensionless amplitude of the EM field $a_0$
\begin{equation}
    a_0  = \frac{eE}{m c \omega},
\end{equation}
where $m$ and $e>0$ are the electron mass and charge respectively and $\omega$ is the characteristic frequency of the EM field.
In the regime $a_0 \gg 1$ the characteristic radiation formation length $\lambda_f$ in most cases can be estimated as $\lambda / a_0 \ll \lambda$ where $\lambda = 2\pi c/\omega$, i.e. individual acts of radiation occur almost instantaneously compared to the variation scale of EM field and thus on a radiation formation length EM fields can be assumed constant.
In this locally constant field approximation (LCFA~\cite{nikishov1964quantum, berestetskii1982quantum, ritus1985quantum}) total radiation probabilities and emission spectrum shapes depend only on the QED parameter $\chi$
\begin{equation}
    \chi = \frac{\gamma}{E_S} \sqrt{{\left( \vb{E} + \vb{v}\times\vb{B} \right)}^2 - {\left(\vb{v}\vb{E} \right)}^2} ,
\end{equation}
where $\gamma$ and $\vb{v}$ are the electron Lorentz-factor and velocity normalized to $c$ respectively, $\vb{E}$ and $\vb{B}$ are the electric and the magnetic field respectively, $E_S = m^2 c^3/e\hbar$ is the critical Sauter-Schwinger field~\cite{berestetskii1982quantum}, $\hbar$ is the Planck constant.
These probabilities can be calculated analytically in either classical ($\chi \ll 1$) or quantum ($\chi \gg 1$) regimes
\begin{equation}
    W_{rad} \approx \alpha \frac{m c^2}{\gamma\hbar} \times
    \begin{cases}
        1.4 \chi, & \chi \ll 1, \\
        0.7 \chi^{2/3}, & \chi \gg 1,
    \end{cases}
\end{equation}
where $\alpha = e^2/\hbar c$ is the fine structure constant.
Note that different approaches has been proposed for calculation of radiation probabilities when LCFA breaks~\cite{khokonov2002standard, ilderton2019extended, heinzl2020locally, gelfer2022nonlinear, Podszus19}.

In LCFA the characteristic distance the ultrarelativistic electron travels between two consecutive photon emission $\lambda_W$ can be estimated as $c/W_{rad}$, which in both classical and quantum cases is at least $1/\alpha\approx137$ times longer than the radiation formation length.
Note, however, that the above-mentioned estimate for the radiation formation length actually depends on the frequency of the emitted radiation~\cite{Artemenko20} and can be inaccurate for $\chi \gtrsim 10$.
Furthermore, the ratio between mean free path $\lambda_W$ and EM field wavelength can be estimated as follows
\begin{equation}
    \frac{\lambda_W}{\lambda} \approx \frac{1}{\alpha a_0} \times
    \begin{cases}
        1, & \chi \ll 1, \\
        \chi^{1/3}, & \chi \gg 1.
    \end{cases}
\end{equation}
So, as $\chi \lesssim 10$ for most experiments of the nearest future, and for $a_0 \gg 137$ the hierarchy of the characteristic scales of the problem is the following
\begin{equation}
    \lambda_f \ll \lambda_W \ll \lambda,
\end{equation}
which means that the electron moves classically between short but frequent acts of the photon emission.

In that case we can approximate the effect of the radiation recoil as an additional continuous force acting on a particle, i.e. the motion equations take form
\begin{align}
    \label{eq.dpdt}
    & \dv{\vb{p}}{t} = -\vb{E} - \vb{v}\times\vb{B} - F_{rr}\vb{v} ,\\
    \label{eq.dgdt}
    & \dv{\gamma}{t} = -\vb{v}\vb{E} - F_{rr} v^2 ,
\end{align}
where electron momentum $\vb{p}$ is normalized to $m c$, time $t$ is normalized to $1/\omega$, electric and magnetic fields are normalized to $m c \omega / e$, $F_{rr}$ is the total radiation power normalized to $mc^2\omega$ given by the expression
\begin{equation}
    F_{rr} = \frac{\alpha a_S}{3 \sqrt{3}\pi}
    \int_{0}^{\infty}\frac{4u^{3}+5u^2+4u}{(1+u)^{4}}K_{2/3}\left(\frac{2u}
    {3\chi}\right) \dd u.
\end{equation}
In the limiting cases the above expression simplifies 
\begin{equation}
    F_{rr} \approx \alpha a_S \times
    \begin{cases}
        0.67 \chi^2, & \chi \ll 1, \\
        0.37 \chi^{2/3}, & \chi \gg 1.
    \end{cases}
\end{equation}
This approach to description of the electron dynamics with account of radiation reaction is commonly referred to as semi-classical~\cite{kirk2009pair,bulanov2013electromagnetic,esirkepov2014attractors, niel2018quantum, gonoskov2021charged}.
In the quantum regime radiated photon can carry away a significant portion of the electron energy and as these radiations are stochastic, electrons contained in a small phase volume can significantly diverge in the phasespace after some time.
Eqs.~\eqref{eq.dpdt}--\eqref{eq.dgdt} essentially describe the 0-th moment, i.e. trajectory of the center of mass of the electron distribution function, while effects caused by the probabilistic nature of radiation, such as straggling and quenching~\cite{shen1972energy, duclous2010monte, harvey2017quantum, gonoskov2021charged}, lead to diffusion of the distribution function and thus cannot be captured using this approach. 
In that case a more accurate description involves equations for higher moments of the distribution function.
This approach was applied to calculate mean particle energy and energy spread in different fields configurations in Refs.~\onlinecite{neitz2013stochasticity, ridgers2017signatures, niel2018quantum}.
If on the contrary $\chi \lesssim 1$ then recoil from a single photon radiation is small and approximation of continuous recoil is sufficient to describe the electron dynamics.

Another important consideration in research of effect of radiation reaction is its dependence on the internal degree of freedom of the electron~---~spin.
Strictly speaking quasi-classical limit of the Dirac equation leads to motion equations where both orbital motion of the electron and evolution of its spin are coupled.
In particular, one should add the Stern-Gerlach force~\cite{gerlach1922experimentelle} in the equation for the electron momentum and describe spin dynamics via Thomas-Bargmann-Michel-Telegdi (T-BMT)~\cite{thomas1926motion, bargmann1959precession} equation.
Note that although the latter is strictly valid in homogeneous EM fields, it still can be used in heterogeneous fields if Stern-Gerlach force can be neglected~\cite{mane2005spin}.
One can estimate that the ratio between the Lorentz force and Stern-Gerlach force is of the order of $\hbar\omega/mc^2$, thus for optical frequencies the latter can be neglected with a large margin of accuracy.
In that case spin dynamics is decoupled from the electron orbital motion and can be calculated after the electron trajectory is found.
Radiation reaction can again couple spin dynamics and electron orbital motion, since radiation probabilities depend on spin of the electron (and polarization of the emitted photon).
Note that an order of magnitude estimates made above where radiation probabilities are averaged over initial and summed over final polarization states of the electron remain valid.
Although in a certain scenarios assuming that electrons are generally not polarized can be invalid, since radiation probabilities of spin up and spin down electron are different.
Resolving electron polarization can lead to effects such as significant increase of pair production during QED cascade development~\cite{Seipt2021PolarizedQEDcascad}, production of polarized high-energy particles~\cite{wen2019polarized, li2019ultrarelativistic}, spatially-inhomogeneous polarization~\cite{gong2021retrieving}, etc.
In this paper, such effects caused by spin dynamics are not covered.

Previous studies have shown that the problem can be simplified even further.
In particular, to some extent radiation reaction can be accounted for implicitly, i.e.~without specifying the expression for radiation reaction in motion equations~\cite{samsonov2018,gonoskov2018radiation}.
This is done by noticing that in constant homogeneous EM fields electron motion is stable if the electron does not experience transverse acceleration.
As radiation probability is essentially proportional to the transverse acceleration, direction of such motion is called radiation-free (RFD). 
As this direction corresponds to vanishing of transverse acceleration and radiation recoil is directed against the electron velocity, to find RFD one does not need to specify expression for radiation probability at all.
Timescale $\tau_v$ on which the electron velocity approaches RFD in constant fields is of the order of $\gamma mc/eE$.
If EM fields are varying with characteristic frequency $\omega$ then radiation-free direction defined by local and instant field configuration changes at the same timescale.
Without radiation reaction one can estimate that $\gamma \sim a_0$, thus by the time an electron velocity approaches RFD, the latter itself changes, so geometric relation between the electron velocity and RFD is arbitrary.
This is not the case in the strongly radiation-dominated regime, though, when by definition $\gamma \ll a_0$ and thus EM field orients the electron velocity much faster than the field changes itself, so the electron velocity quickly aligns to the RFD defined by local and instant electric and magnetic fields.
So to approximately determine electron trajectory one can assume that at each time instant the electron velocity coincides with RFD.
While this approach allows to describe dynamics of the electron in the strongly radiation-dominated regime without specifying expression for the radiation power, it is quite limited for a couple of reasons.
First, electron velocity converges to RFD fast enough only at extremely large intensities $I \gtrsim \SI{e25}{W/cm^2}$.
And second, this approach does not allow one to find the electron energy and radiation losses while approaching RFD as particle energy is assumed to be indefinitely large albeit much smaller than the field amplitude at the same time.
Despite its apparent drawbacks, this approach was recently successfully applied for describing electron motion in an astrophysical environment~\cite{jerome2022particle}.
In this paper, we advance this radiation-free approach to overcome its inherent problems and to describe dynamics of the electron in the strongly radiation-dominated regime more precisely.

The paper is organized as follows.
In Sec.~\ref{sec.RFD} we reintroduce the concept of radiation free-dynamics and extend it by application of perturbation theory.
In Sec.~\ref{sec.Examples} we consider several EM field configurations where obtained reduced motion equations can be explicitly solved.
In Sec.~\ref{sec.Conclusion} we discuss domain of applicability of the developed approach and draw conclusions.

\section{Radiation-free approach}
\label{sec.RFD}

Let us start by introducing a radiation-free approach to description of the electron dynamics, loosely following original papers~\cite{samsonov2018,gonoskov2018radiation}.
The equations governing the electron dynamics in EM field with account of radiation reaction can be written in terms of its velocity $\vb{v}$ and Lorentz-factor $\gamma$
\begin{align}
    \label{eq.base1}
    &\dv{\gamma}{t} = -\vb{vE} - F_{rr} v^2 , \\
    \label{eq.base2}
    &\dv{\vb{v}}{t} = - \frac{1}{\gamma}\left( \vb{E + v\times B - v\left( vE \right)} + \frac{F_{rr}\vb{v}}{\gamma^2} \right) ,
\end{align}
Since only for ultrarelativistic particle ($\gamma \gg 1$) the radiation reaction is sufficient to alter the particle dynamics, the last term in Eq.~\eqref{eq.base2} can be omitted.
There exist a formal stationary solution $\vb{v}_0$ of these equations, corresponding to vanishing of the transverse force acting on the electron and in turn vanishing of radiation reaction.
Because of that property this solution is called radiation-free direction (RFD).
\begin{equation}
    \label{eq.rfd1}
    \vb{E} + \vb{v}_0\times \vb{B} - \vb{v}_0\left( \vb{v}_0\vb{E} \right) = 0 .
\end{equation}
Note, first, that there always exist a solution to that equation which can be calculated algebraically~\cite{samsonov2018} or geometrically~\cite{gonoskov2018radiation} and, second, that $|\vb{v}_0|=1$ which can be obtained by performing a scalar multiplication of Eq.~\eqref{eq.rfd1} by $\vb{v}_0$, which means that this solution is not entirely physical, i.e. an electron is unable to move in EM field without experiencing transverse acceleration.
To understand how this solution relates to actual solution of the motion equations in the strongly radiation-dominated regime let us consider the following.
By definition in the strongly radiation-dominated regime an electron energy is significantly smaller compared to the energy of a hypothetical electron in the same EM fields but which does not experience radiation reaction.
One can roughly estimate that characteristic energy of the latter electron is of the order of the dimensionless electric field amplitude $E$.
So then for the real electron in the radiation-dominated regime we can estimate $\gamma \ll E$.
Under just made assumption Eq.~\eqref{eq.base2} then states that EM field orients electron velocity much faster than the field itself changes.
So on a timescale of velocity orientation EM fields can be assumed constant and homogeneous.
In that case electron velocity asymptotically tends to RFD.
Neglecting the time it takes for the electron velocity to approach RFD, one can construct an asymptotic trajectory which in some sense serves as an attractor for real electron trajectories
\begin{align}
    \label{eq.rfd2}
    \dv{\vb{r}}{t} = \vb{v}_0\left(\vb{E}(\vb{r},t), \vb{B}(\vb{r},t)\right).
\end{align}
As mentioned in~\hyperref[sec.Intro]{Introduction}, although Eqs.~\eqref{eq.rfd1}--\eqref{eq.rfd2} describe electron dynamics in the strongly-radiation dominated regime with radiation reaction being accounted for implicitly, there are two drawbacks of this approach.
First, these asymptotic trajectories describe real particle trajectories well only at extremely large intensities $I \gtrsim \SI{e25}{W/cm^2}$.
This comes from the fact that for most field configurations characteristic time at which the electron velocity approaches RFD is underestimated in above reasoning.
And second, this approach does not allow one to find the electron energy and radiation losses while approaching this asymptotic trajectory as particle energy assumed to be indefinitely large (albeit much smaller than the field amplitude at the same time).

To get rid of the mentioned issues we develop a perturbation theory, assuming that electron velocity deviates from radiation-free direction and this deviation is small, i.e.
\begin{equation}
    \vb{v}=\left( 1 - \frac{\delta^2}{2} \right) \vb{v}_0 + \vb{v}_1,
\end{equation}
where $\vb{v}_1\perp\vb{v}_0$ and $\delta$ can be found from the condition that $|\vb{v}|^2= 1-\gamma^{-2}$, from which we get 
\begin{equation}
    \delta^2\approx v_1^2+\gamma^{-2}.
\end{equation}
Substituting this into Eqs.~\eqref{eq.base1}--\eqref{eq.base2}, utilizing Eq.~\eqref{eq.rfd1} and keeping only terms of the first order on $v_1$~(see expansion up to the terms of the second order in Appendix~\ref{app.Terms}), we obtain a set of general equations governing electron dynamics in the strongly radiation-dominated regime
\begin{align}
    \label{eq.main1}
    &\dv{\vb{v}_1}{t}=\frac{\vb{F}_1}{\gamma} - \dv{\vb{v}_0}{t} - \vb{v}_0\left( \vb{v}_1\dv{\vb{v}_0}{t} \right) ,        \\
    \label{eq.main2}
    &\dv{\gamma}{t} = -\vb{v}_0\vb{ E} \left( 1 - \frac{v_1^2}{2} - \frac{1}{2\gamma^2} \right) - \vb{v}_1 \vb{E} - F_{rr}(\chi) ,                                                                       \\
    \label{eq.f1}
    &\vb{F}_1 = (\vb{v}_0\vb{ E})\vb{v}_1 + (\vb{v}_0\vb{ B})[\vb{v}_0\times\vb{v}_1] + \mathcal{O}\left( \delta^2 \right), \\
    \label{eq.base_chi}
    &\chi = \frac{\gamma |\vb{F}_1| }{a_S}.
\end{align}
Note that although $\chi$ is proportional to a small term $v_1$ it can be arbitrarily large due to the factor $\gamma$, therefore the term $F_{rr}$ should be kept in all expansion orders, which leads to motion equations remaining non-linear.
The full time derivatives should be considered as derivatives of the vector field $\vb{v}_0$ along the real electron trajectory $\vb{r}(t)$, i.e.
\begin{equation}
    \dv{\vb{v}_0}{t} = \pdv{\vb{v}_0}{t} + (\vb{v},\nabla)\vb{v}_0 .
\end{equation}
Let us consider equation for the magnitude of the vector $\vb{v}_1$
\begin{equation}
    \label{eq.dv2dt}
    \frac{1}{2}\dv{v_1^2}{t} = - \vb{v}_1\dv{\vb{v}_0}{t} + \frac{v_1^2(\vb{v}_0 \vb{E})}{\gamma}.
\end{equation}
In constant EM fields ($\dd\vb{v}_0/\dd t = 0$) one can estimate characteristic timescale at which the electron velocity approaches RFD
\begin{equation}
    \tau_v = \frac{\gamma}{|\vb{v}_0 \vb{E}|} \sim \frac{\gamma}{a_0}.
\end{equation}
But for varying EM fields the sign of the first term in Eq.~\eqref{eq.dv2dt} can be arbitrary and its magnitude as large as $v_1$, so condition $\gamma \ll a_0$ alone is not enough to justify description of the electron dynamics with Eq.~\eqref{eq.rfd2} in an arbitrary field configuration.
Instead on should use set of Eqs.~\eqref{eq.main1}--\eqref{eq.main2}, where variation of the RFD is taken into account.
Moreover these equations allow to find electron energy and radiative losses.

Explanation of the performed procedure to obtain reduced motion equations can be done in few simple steps.
First, it is shown that there exist a preferred radiation-free direction, to which an electron velocity approaches in constant fields.
By decomposing electron velocity in a new basis where one axis coincides with RFD motion equations can be split.
Motion along RFD is essentially described via the particle energy, while equations for transverse velocity can be expanded into a series, which evidently converges since magnitude of the velocity being strictly smaller than unity.
Although final set of equations still remain non-linear and cannot be solved explicitly in an arbitrary field configuration, examples considered below show that this approach can be superior to solving non-reduced Newton equations.

\section{Example problems}
\label{sec.Examples}
Below we consider several exemplary field configurations in which the obtained equations~\eqref{eq.main1}--\eqref{eq.main2} can be solved explicitly.

\subsection{Generalized Zeldovich problem}
\label{sub.Zeldovich}

\begin{figure}
    \includegraphics[width=85mm]{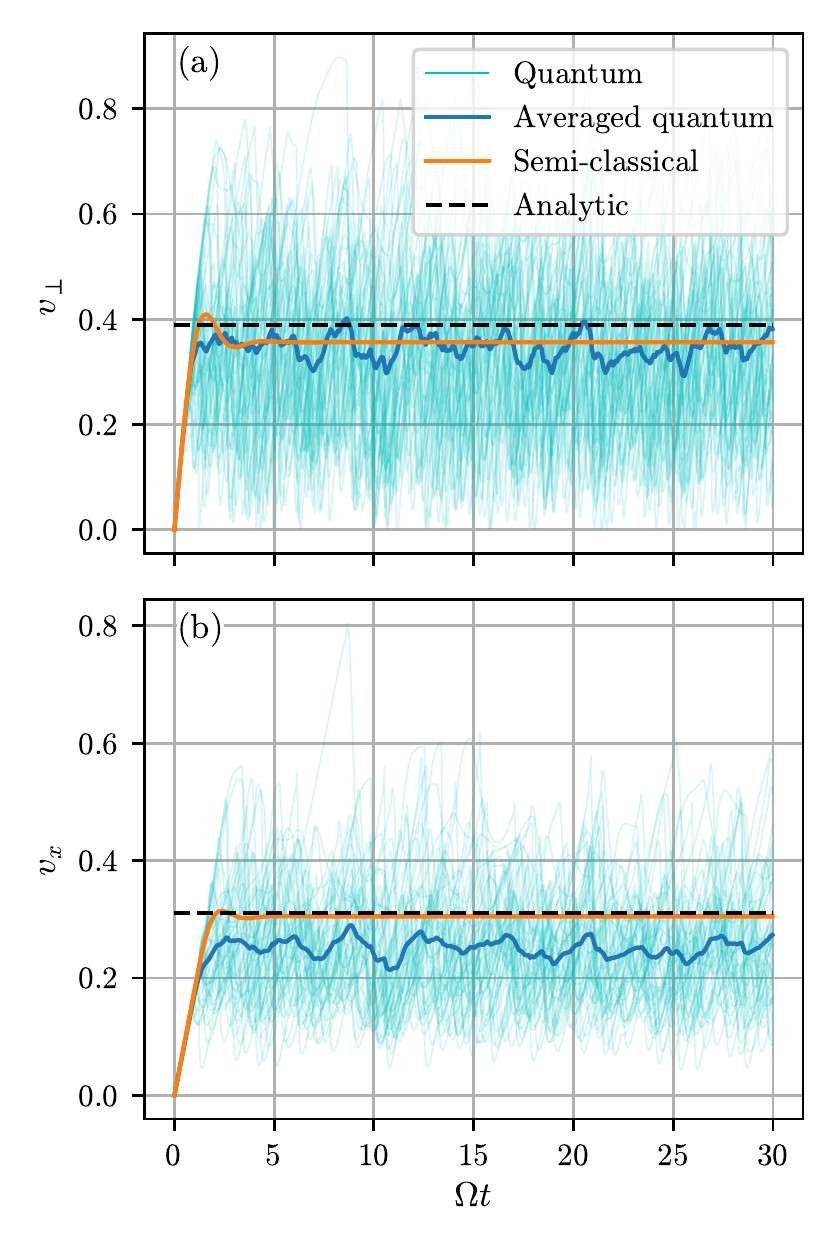}
    \caption{Electron dynamics in an electric field with dimensionless amplitude $eE/mc\Omega = \num{2500}$ and parallel magnetic field with dimensionless amplitude $eB/mc\Omega = \num{2000}$ rotating with the angular frequency $\Omega$, corresponding to the wavelength $\lambda= \SI{1}{\um}$: (a) component of the electron velocity transverse to electric field, (b) component of the electron velocity along the angular velocity vector $\vb{\Omega}$. Orange (cyan) lines correspond to numerical solution of non-reduced motion equations~\eqref{eq.base1}--\eqref{eq.base2} with account of radiation reaction via semi-classical (quantum) approach, blue lines correspond to the average value of 100 `quantum' solutions, Black dashed lines correspond to the analytical solution~\eqref{eq.Zeld_sol1}--\eqref{eq.Zeld_sol2}.}
    \label{fig.zeldovich}
\end{figure}

The equations of electron motion with radiation reaction force can be integrated analytically for the rotating uniform electric field which has been first demonstrated by Zeldovich~\cite{Zeldovich75}. Recently the Zeldovich's solution has been extended to the configuration of the rotating electric and magnetic field which are parallel to each other~\cite{kostyukov2016production}. Let us analyze the last configuration within our approach. We assume that the electric and magnetic fields are uniform, parallel, and rotate with velocity $\vb{\Omega}$.
The radiation-free direction in this configuration is anti-parallel to the electric field: $\vb{v}_0~=~-\vb{E}/E~=~-\vb{e}$.
Let us consider a stationary solution when the deviation vector $\vb{v}_1$ rotates synchronously with electric and magnetic fields.
In that case all the time derivatives can be replaced with cross product $\vb{\Omega}\times$.
Eq.~\eqref{eq.main1} is then written as follows
\begin{equation}
    \label{eq.zeldovich}
    \vb{\Omega}\times\vb{v}_1 = -\frac{E}{\gamma}\vb{v}_1 + \frac{B}{\gamma}\vb{e\times v}_1 + \vb{\Omega\times e} + v_1 \vb{e}.
\end{equation}
Keeping in mind that $\vb{v}_1 \perp \vb{v}_0$ we can express $\vb{v}_1$ the following way
\begin{equation}
    \vb{v}_1 = v_\perp \vb{\Omega}\times\vb{e} + v_x \vb{\Omega} .
\end{equation}
In that case Eq.~\eqref{eq.zeldovich} splits into a set of linear equations which solutions can be easily found
\begin{align}
    \label{eq.Zeld_sol1}
    & v_x = \frac{\gamma B}{E^2 + B^2} ,\\
    \label{eq.Zeld_sol2}
    & v_\perp = \frac{\gamma E}{E^2 + B^2}.
\end{align}
This stationary solution corresponds to constant radiative losses and constant energy.
That is why final relation between the electron energy and fields can be obtained from Eq.~\eqref{eq.main2}
\begin{equation}
    E = F_{rr}(\chi) = F_{rr}\left(\frac{\gamma^2}{a_S}\right) .
\end{equation}
This result coincides exactly with the result obtained in Ref.~\onlinecite{kostyukov2016production} and in the special case $B=0$~---~to the original Zeldovich's solution~\cite{Zeldovich75}. Comparison of the obtained solution with numerical solution of non-reduced motion equations~\eqref{eq.base1}--\eqref{eq.base2} in both semi-classical and quantum approach to radiation reaction is presented in Fig.~\ref{fig.zeldovich}.
Magnitude of the EM field was chosen in such a way that average value of $\chi$ parameter of an electron is around 5.
This was done to show that our approach describes `average' electron well, although parameters of individual electrons with the same initial conditions spread quite significantly dues to stochastic nature of radiation in quantum regime, as mentioned in~\hyperref[sec.Intro]{Introduction}.

\subsection{Monochromatic linearly polarized plane wave}
\label{sub.PlaneWave}

\begin{figure}\centering
    \includegraphics[width=85mm]{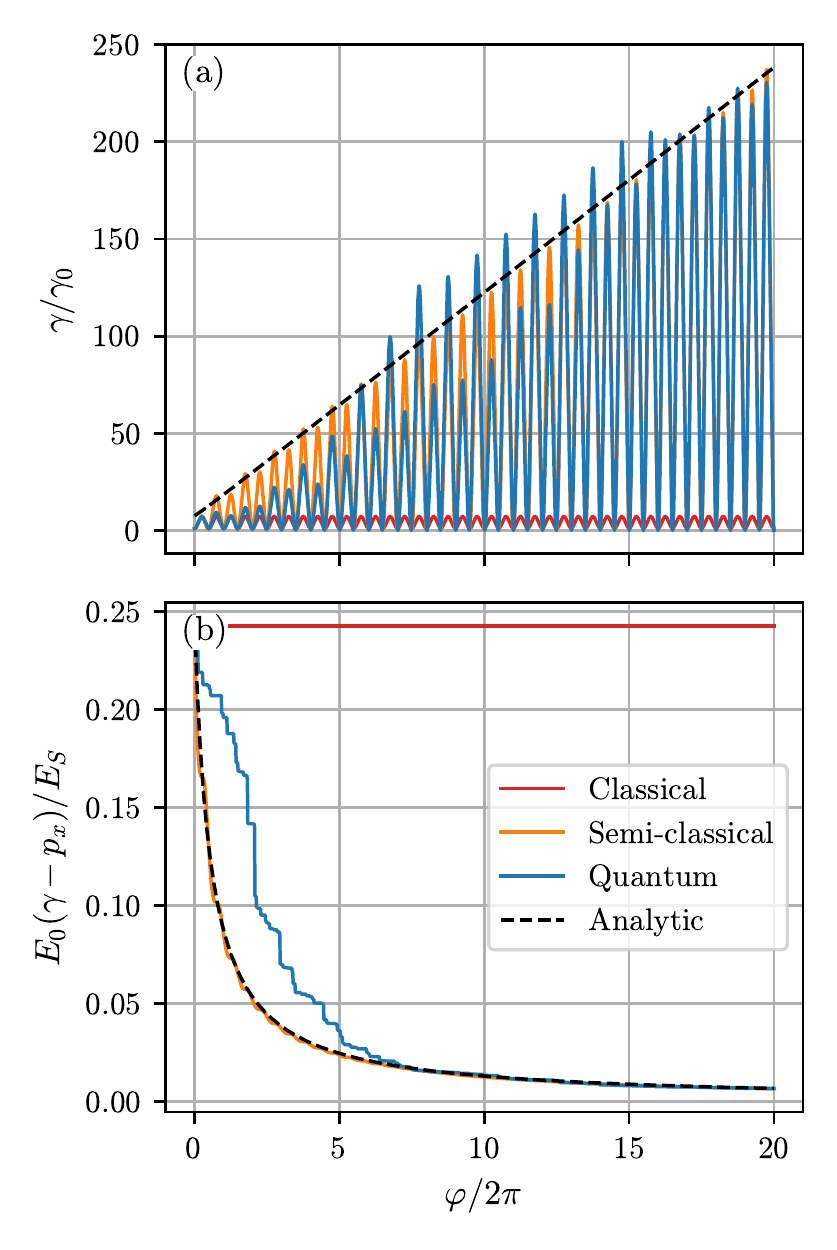}
    \caption{Dynamics of an electron with initial momentum $p_x = -100\ m c$ in a plane wave with amplitude $E_0 = 500$ and wavelength $\lambda = \SI{1}{\um}$ propagating along the $x$-axis: (a) energy of the electron normalized to its initial value, (b) maximum value of the QED parameter $\chi$: $E_0(\gamma~-~p_x)/E_S$. Red lines correspond to classical solution without radiation reaction, orange (blue) lines~---~to numerical solution of non-reduced motion equations~\eqref{eq.base1}--\eqref{eq.base2} with account of radiation reaction via semi-classical (quantum) approach. Black dashed lines correspond to analytical solution~\eqref{eq.fin_pw1}--\eqref{eq.fin_pw2}.}
    \label{fig.pw}
\end{figure}

Some interesting results can be obtained if one applies our approach to the electron motion in a plane wave.
In this configuration radiation-free direction coincides with the direction of the Poynting vector
\begin{equation}
    \vb{v}_0=\frac{\vb{E\times B}}{E^2}.
\end{equation}
where both $\vb{E}$ and $\vb{B}$ are functions of the phase $\varphi=x-t$.
Note that the fact that a strong wave pushes the electron in the direction of its propagation was recently observed in an exact analytical solution of the electron motion in a plane wave~\cite{ekman2021exact}.
For the sake of simplicity let us assume that initially the deviation vector $\vb{v}_1$ is parallel to the electric field, in which case it will stay so at any time instance.
It is also more convenient to write down equations in terms of momentum $\vb{p}_1 = \gamma \vb{v}_1$ and phase $\varphi$.
Since $\chi$ oscillates with a constant amplitude in a plane wave without account of radiation reaction and radiation leads only to decrease of $\chi$, eventually any electron will reach classical regime when $\chi \ll 1$, so we will only consider electron dynamics in that regime.
In that case Eqs.~\eqref{eq.main1}--\eqref{eq.main2} take form
\begin{align}
    \label{eq.pw1}
    & \dv{\gamma}{\varphi} = -2\frac{p\gamma E}{1 + p^2 } - 2 A_{rr} E^2 \left( 1 + p^2 \right) ,\\
    \label{eq.pw2}
    & \dv{p}{\varphi} = -E - 2 A_{rr} E^2 \frac{p}{\gamma}\left( 1 + p^2 \right).
\end{align}
where $A_{rr} = \alpha/6E_S$.
Note that in a plane-wave configuration $\vb{F}_1$ defined in~\eqref{eq.f1} contains only the terms of the second and higher orders of smallness on $v_1$, derivation of which can be found in Appendix~\ref{app.Terms}.
Let us start solving Eqs.~\eqref{eq.pw1}--\eqref{eq.pw2} by assuming $A_{rr}=0$ and $E=E_0\cos\varphi$.
In that case it is easy to obtain the following solution
\begin{align}
    \label{eq.pw_sol1}
    & p_{pw} = -E_0\sin\varphi ,\\
    \label{eq.pw_sol2}
    & \gamma_{pw} = \gamma_0 \left( 1 + {p_{pw}}^2 \right),
\end{align}
where it was assumed that initially the electron momentum has only component along the direction of the plane wave propagation.
Note that the exact solution of the electron motion equations in a linearly-polarized plane wave (see e.g. Ref.~\onlinecite{landau2013classical}) exactly coincide with solution~\eqref{eq.pw_sol1}--\eqref{eq.pw_sol2} in the limit $\gamma_0 \gg 1$.
So our method, which is essentially a series expansion, can be applied even for certain problems where radiation reaction is not accounted.

When $A_{rr} \neq 0$ it can be shown from Eqs.~\eqref{eq.pw1}--\eqref{eq.pw2} that there is an asymmetry in the particle motion in accelerating and decelerating phases.
This leads to a non-zero energy gain on a single period. 
Applying several additional assumptions we can obtain the following solution (see detailed step-by-step solution in Appendix~\ref{app.PW})
\begin{align}
    \label{eq.fin_pw1}
    & \gamma \approx \gamma_0 \left( 1 + E_0^2\sin^2\varphi \right) \left( 1 + \frac{A_{rr}E_0^2}{\gamma_0}\varphi \right) , \\
    \label{eq.fin_pw2}
    & v \approx - \frac{E_0\sin\varphi}{\gamma} .
\end{align}
Rewriting the solution in terms of lab time $t$ yields the following result
\begin{align}
    \label{eq.pw_lab_time}
    \langle \chi \rangle,\ \langle v_1 \rangle,\ \langle \gamma^{-1} \rangle \propto \left(A_{rr} t \right)^{-1/3}.
\end{align}
Note that average behavior of the dependencies~\eqref{eq.pw_lab_time} asymptotically coincides with the one extracted from the exact solution derived in Ref.~\onlinecite{di2008exact}.

In obtaining the above solution we assumed that at the initial moment transverse momentum of the particle is equal to zero and longitudinal momentum is sufficiently large and positive.
To obtain a solution with an arbitrary initial conditions one can perform a Lorentz-boost to an auxiliary reference frame where above assumptions are satisfied and then transform the above solution back to an initial reference frame.
Solution shown in Fig.~\ref{fig.pw} was obtained in such a way, where an auxiliary reference frame is moving with the velocity corresponding to Lorentz-factor 1000 along the negative $x$-axis in the laboratory reference frame, so in the auxiliary reference frame the initial longitudinal electron momentum is approximately equal to 5~$m c$.

The obtained solution is not only non-periodic but also features quite an unexpected behavior: instead of slowing down the electron, radiation reaction actually allows the electron to gain infinite energy (for infinite time obviously).
Although this behavior has been reported before~\cite{gunn1971motion, grewing1973acceleration, thielheim1993particle} and is confirmed by numerical solution of non-reduced motion equations~\eqref{eq.base1}--\eqref{eq.base2} (see Fig.~\ref{fig.pw}), it does not appear to be widely acknowledged.
A simple reasoning can explain this seemingly controversial phenomenon.
For that it is more convenient to resort to quantum description of radiation reaction.
In a relativistically strong plane wave ($E \gg 1$) formation length  of the radiation can be estimated as $\lambda/E \ll \lambda$, which can be interpreted that an electron moves classically between short acts of photon emission.
Without radiation reaction quantity $\gamma - p_x$ is the constant of motion, where $p_x$ is the electron momentum along the direction of the plane wave propagation (see red line in Fig.~\ref{fig.pw}~(b)).
The radiation probability depends on a QED parameter $\chi$ which in the plane wave configuration is written as follows
\begin{equation}
    \chi = \frac{E(\varphi)}{E_S} \left( \gamma - p_x \right) .
\end{equation}
As radiation formation length is much smaller than the wavelength the fields can be assumed constant and from energy and momentum conservation laws during the photon emission it follows that the parameter $\chi$ of the electron strictly reduces after the emission (see distinct jumps corresponding to radiation of individual photons in blue line in Fig.~\ref{fig.pw}~(b)).
So we can conclude that due to radiation reaction the quantity $\gamma - p_x$ asymptotically tends to zero which can be satisfied only when $p_x$ (and $\gamma$ correspondingly) grows indefinitely.

\subsection{Plasma accelerator}
\label{sub.Accelerator}
Finally, let us consider a toy model of a plasma accelerator and derive conditions of a known stable solution in a radiation-dominated regime~\cite{kostyukov2012radiative, golovanov2021radiation} using our approach.
For this we assume that EM field consists of a uniform accelerating field $\vb{z}_0 E_{acc}$ and a linear focusing field $\vb{y} E_{foc}$ in which the electron undergoes betatron oscillations.
To find a solution which corresponds to the averaged radiation losses being constant on many betatron periods we will assume that any function of the QED parameter $\chi$ is a strictly periodic function of time, thus average of any function of $\chi$ is also constant, i.e.
\begin{equation}
    \label{eq.acc_const_av}
    \dv{\langle \chi^2 \rangle}{t} = 0 ,
\end{equation}
where $\chi^2$ is used for convenience.
In the considered field configuration from Eq.~\eqref{eq.base_chi} we get
\begin{equation}
    \chi = \frac{\gamma E v_1}{a_S}.
\end{equation}
Hereinafter we will assume that $\gamma \approx \langle \gamma \rangle$, i.e. oscillations of the particle energy are much smaller than energy itself, so we can extract $\gamma$ from angular brackets.
Also as particle is accelerated $\gamma$ grows with time, while $\chi$ remains constant on average.
This means that oscillation amplitude of $v_1$ decreases, so at later times we can safely assume that the electric field experienced by the electron is mostly accelerating, so $E \approx E_{acc}=\text{const}$.
\begin{figure}
    \includegraphics[width=85mm]{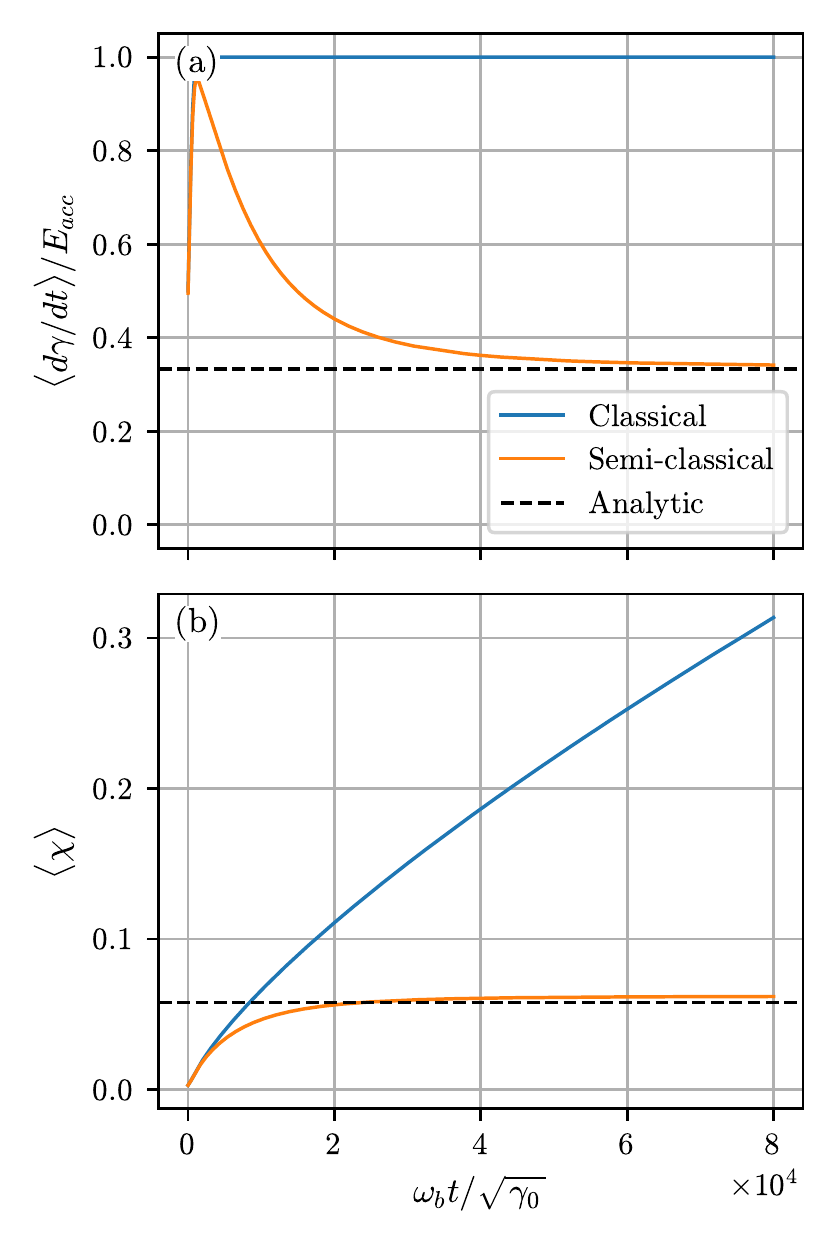}
    \caption{Electron dynamics in a model accelerator with $E_{acc} = \SI{30}{TV/m}$ and $E_{foc}$ growing linearly from $0$ to $\SI{30}{TV/m}$ at displacement $\SI{0.1}{\um}$: (a) average accelerating rate, (b) average value of QED parameter $\chi$. Time is normalized to initial value of inverse betatron frequency $\omega_b/\sqrt{\gamma_0}$ of the electron. Blue lines correspond to the solution of non-reduced motion equations~\eqref{eq.base1}--\eqref{eq.base2} without account of radiation reaction, orange lines~---~to the solution with radiation reaction accounted via a semi-classical approach. Black dashed line corresponds to analytical solution~\eqref{eq.acc_23}.}
    \label{fig.accelerator}
\end{figure}
Both these assumption are reliably confirmed by numerical simulations and remain valid in the final solution we obtain below.
Utilizing these assumptions and expanding Eq.~\eqref{eq.acc_const_av} yields the following
\begin{align}
    \label{eq.acc_first_equity}
    \left\langle F_{rr} \right\rangle \left\langle v_1^2 \right\rangle + \gamma \left\langle \vb{v}_1 \dv{\vb{v}_0}{t} \right\rangle = 0.
\end{align}
If we differentiate this expression again we obtain the following result
\begin{equation}
    \label{eq.acc.interm}
    \begin{multlined}[c][7cm]
         \frac{\langle F_{rr} \rangle }{\gamma} (3 \langle F_{rr} \rangle - 2 E_{acc}) \left\langle v_1^2  \right\rangle + \\
         + \gamma\left\langle \vb{v}_1\dv[2]{\vb{v}_0}{t} - \left( \dv{\vb{v}_0}{t} \right)^2 \right\rangle = 0 .
    \end{multlined}
\end{equation}
To deal with the last two terms in the above expression, let us write the equation for the electron trajectory
\begin{equation}
    \label{eq.acc.r}
    \dv{\vb{r}}{t} = \vb{v}_0 + \vb{v}_1, 
\end{equation}
where
\begin{equation}
    \vb{v}_0 = -\frac{-\vb{z}_0 E_{acc} + \vb{y} E_{foc}}{E} \approx \vb{z}_0 - \vb{y} \frac{E_{foc}}{E_{acc}} \equiv \vb{z}_0 - \kappa\vb{y}.
\end{equation}
If we assume that betatron oscillations are harmonic, i.e.
\begin{equation}
    y = y_0\cos\omega t,
\end{equation}
then from the $y$-component of the Eq.~\eqref{eq.acc.r} we get
\begin{equation}
    v_{1,y} = y_0 (\kappa\cos\omega t - \omega\sin\omega t).
\end{equation}
To calculate the average of last two terms in Eq.~\eqref{eq.acc.interm} note that $\dd\vb{v}_0/\dd t = -\kappa \dd\vb{y}/\dd t$
\begin{equation}
    \begin{aligned}
        &\left\langle \vb{v}_1\dv[2]{\vb{v}_0}{t} - \left( \dv{\vb{v}_0}{t} \right)^2 \right\rangle = y_0^2 \omega^2\kappa\times \\
        &\times\left(\kappa\left\langle  \cos^2\omega t -  \sin^2\omega t \right\rangle - \omega \left\langle\sin\omega t \cos\omega t\right\rangle\right)= 0.
    \end{aligned}
\end{equation}
So finally in a model accelerator we get
\begin{equation}
    \label{eq.acc_23}
    \langle F_{rr} \rangle = \frac{2}{3} E_{acc} .
\end{equation}
So on average the particle is accelerated only at a third of the classical accelerating rate~\cite{kostyukov2012radiative, golovanov2021radiation}.
Fig.~\ref{fig.accelerator} shows that obtained solution coincides with numerical solution of non-reduced motion equations~\eqref{eq.base1}--\eqref{eq.base2} quite well.
It should be noted that we did not utilize a specific expression for power of radiative losses, though a more rigorous derivation of the relation~\eqref{eq.acc_23} shows that the result actually depends on the scaling law of the radiation power on parameter $\chi$, e.g. according to Ref.~\onlinecite{golovanov2021radiation} in fully quantum regime when $\chi\gg1$ and $F_{rr}\propto \chi^{2/3}$ the relation~\eqref{eq.acc_23} should in fact be slightly different, in particular
\begin{equation}
    \langle F_{rr} \rangle = \frac{12}{19} E_{acc} ,
\end{equation}
which is only 5\% different from~\eqref{eq.acc_23}.
Reasoning using our approach cannot exactly reproduce this minor difference though, due to approximations used throughout, in particular neglecting terms proportional to $1/\gamma^2$ when calculating $\chi$.

\section{Discussion and conclusion}
\label{sec.Conclusion}

In this paper we have developed an approach to tackle the problem of a single electron dynamics in and arbitrary EM field in the strongly radiation-dominated regime.
It is shown that electron velocity approaches a certain direction, moving along which an electron does not experience transverse acceleration and thus does not radiate.
If we assume that electron velocity slightly deviates from this radiation-free direction then motion equations can be simplified.
In certain EM field configurations such simplification is enough to allow an analytical solution of the electron motion equations.
Remarkably, in a plane wave example the obtained solution is valid even if radiation reaction is not accounted.
This shows that the domain of the applicability of our method is larger than was initially expected.
This can be partially attributed to the fact that our approach is based on an expansion of motion equations written in terms of the electron velocity.
Since magnitude of this vector is smaller than unity, series expansion in terms of velocity should converge.
The rate at which these series converge depend on how close the zeroth order is to the real value.
We have shown that in the strongly radiation-dominated regime radiation-free direction can be chosen as a zeroth order approximation for the direction of the electron velocity.
But plane wave example shows that the same expansion can be valid even without account of radiation reaction in certain field configurations.

It should be noted that the developed approach is valid when continuous radiation recoil approximation is justified.
This approximation is mostly determined by the value of QED parameter $\chi$.
In particular, in sufficiently quantum regime, when $\chi\gg 1$, electron dynamics can become stochastic and thus electron distribution function can evolve in a  complex way.
In that case equations for higher moments of the distribution function can provide a more accurate description, however this lays outside the scope of the paper.

In conclusion, we have proposed a general approach for theoretical investigation of a single particle dynamics in the strongly radiation-dominated regime.
Most importantly the developed method allows to obtain qualitatively new results, compared to the radiation-free approach originally developed in Refs.~\onlinecite{samsonov2018,gonoskov2018radiation}.
We demonstrated application of our method in different EM field configurations.
In particular, we reproduced generalized Zeldovich's solution in rotating parallel electric and magnetic fields~\cite{Zeldovich75, kostyukov2016production}, dampening of an average electron acceleration rate in a model plasma accelerator in the radiation-dominated regime~\cite{kostyukov2012radiative, golovanov2021radiation}, and a peculiar feature of the electron motion in a strong plane wave~---~unlimited longitudinal acceleration~\cite{gunn1971motion, grewing1973acceleration, thielheim1993particle,di2008exact,ekman2021exact}.
Utilizing this approach to explore plasma behavior in a radiation-dominated regime is planned for future.

\begin{acknowledgments}
    This work has been supported by Ministry of Science and Higher Education of the Russian Federation (Agreement No. 075-15-2020-906, Center of Excellence “Center of Photonics”).
    We are grateful to Anton Golovanov for fruitful discussions.
\end{acknowledgments}

\appendix

\section{Deriving reduced motion equations up to the terms of the second order of smallness}
\label{app.Terms}

To obtain reduced motion equations we consider the following representation of the electron velocity
\begin{equation}
    \vb{v} = \vb{v}_0\left(1 - \frac{v_1^2}{2} - \frac{1}{2\gamma^2}\right) + \vb{v}_1.
\end{equation}
Let us substitute this into Eq.~\eqref{eq.base1} and expand it keeping only terms up to the second order, $v_1^2$, and larger
\begin{equation}
    \label{eq.app.2nd_order_A}
    \begin{aligned}
         & \left( 1 - \frac{v_1^2}{2} - \frac{1}{2\gamma^2} \right)\dv{\vb{v}_0}{t} - \frac{\vb{v}_0}{2}\left( \dv{v_1^2}{t}+ \dv{\gamma^{-2}}{t} \right) + \dv{\vb{v}_1}{t} =                                                                                                                                \\
         & = -\frac{1}{\gamma}\left\{ - \left( \frac{v_1^2}{2} + \frac{1}{2\gamma^2} \right) \vb{v}_0 \times \vb{B} + \vb{v}_1 \times \vb{B} + \right. \\
         & \left. +\left( v_1^2 + \frac{1}{\gamma^2} \right) \vb{v}_0(\vb{v}_0 \vb{E}) - \vb{v}_0(\vb{v}_1 \vb{E}) - \vb{v}_1(\vb{v}_0 \vb{E}) - \vb{v}_1(\vb{v}_1 \vb{E})\right\}.
    \end{aligned}
\end{equation}
To eliminate the term $\dd v_1^2/\dd t$ in this equation let us perform a scalar multiplication of this equation on $\vb{v}_1$, keeping in mind that $\vb{v}_1 \vb{v}_0=0$ and neglecting higher-order
\begin{equation}
    \label{eq.app.dv2dt}
    \frac{1}{2} \dv{v_1^2}{t} = - \vb{v}_1\dv{\vb{v}_0}{t} + \frac{v_1^2(\vb{v}_0 \vb{E})}{\gamma}.
\end{equation}
Let us expand the term with $\dd \gamma^{-2} /\dd t$:
\begin{align}
    \label{eq.app.dg2dt}
    \dv{\gamma^{-2}}{t} = -\frac{2}{\gamma^3}\dv{\gamma}{t} \approx \frac{2\vb{v}_0\vb{E} }{\gamma^3}
\end{align}
Substituting~\eqref{eq.app.dv2dt} and \eqref{eq.app.dg2dt} back to Eq.~\eqref{eq.app.2nd_order_A} we get
\begin{align}
    \label{eq.app.2nd_order_B}
    &\begin{multlined}[c][7cm]
         \dv{\vb{v}_1}{t} = -\frac{\vb{F}_1}{\gamma}- \\
         - \left( 1 - \frac{v_1^2}{2} - \frac{1}{2\gamma^2} \right)\dv{\vb{v}_0}{t} -\vb{v}_0\left(\vb{v}_1\dv{\vb{v}_0}{t} \right),
    \end{multlined} \\
    \label{eq.app.f1_A}
    &\begin{multlined}[c][7cm]
        \vb{F}_1 = - \left( \frac{v_1^2}{2} + \frac{1}{2\gamma^2} \right) \vb{v}_0 \times \vb{B} + \vb{v}_1 \times \vb{B} - 
        \\ -\frac{\vb{v}_0(\vb{v}_0\vb{E})}{\gamma^2} - \vb{v}_0(\vb{v}_1 \vb{E}) - \vb{v}_1(\vb{v}_0 \vb{E}) - \vb{v}_1(\vb{v}_1 \vb{E}).
    \end{multlined}
\end{align}
Let us separately examine the vector $\vb{v}_1 \times \vb{B}$:
\begin{equation}
    \label{eq.app.v1b}
    \begin{aligned}
         & \vb{v}_1 \times \vb{B} = -(\vb{v}_0 \vb{B})[\vb{v}_0 \times \vb{v}_1] + \vb{v}_0 (\vb{v}_0[\vb{v}_1\times \vb{B}]).
    \end{aligned}
\end{equation}
Now let us perform a scalar product of the Eq.~\eqref{eq.rfd1} by $\vb{v}_1$
\begin{equation}
    \vb{v}_1 \vb{E} + \vb{v}_1 [\vb{v}_0\times \vb{B}] = 0.
\end{equation}
Performing a cyclic permutation of a triple scalar product we get
\begin{equation}
    \label{eq.app.v0v1B}
    \vb{v}_0 [\vb{v}_1\times \vb{B}] = \vb{v}_1 \vb{E}.
\end{equation}
Substituting this relation into Eq.~\eqref{eq.app.v1b} we obtain
\begin{equation}
    \label{eq.app.v1b_fin}
    \vb{v}_1\times \vb{B} = -(\vb{v}_0 \vb{B})[\vb{v}_0 \times \vb{v}_1] + \vb{v}_0(\vb{v}_1 \vb{E}).
\end{equation}
Finally, substituting this into Eq.~\eqref{eq.app.f1_A} we get
\begin{equation}
    \begin{multlined}[c][7cm]
        \vb{F}_1 = -(\vb{v}_0 \vb{B})[\vb{v}_0 \times \vb{v}_1] - (\vb{v}_0 \vb{E}) \vb{v}_1 - \\
         - \left( \frac{v_1^2}{2} + \frac{1}{2\gamma^2} \right) \vb{v}_0 \times \vb{B} + \frac{\vb{v}_0(\vb{v}_0\vb{E})}{\gamma^2} - (\vb{v}_1 \vb{E})\vb{v}_1. 
    \end{multlined}
\end{equation}

\section{Approximate electron motion in the plane wave}
\label{app.PW}
Let us consider the following plane wave configuration: $\vb{E} = E(\varphi)\vb{y}_0$, $\vb{B} = E(\varphi)\vb{z}_0$, $\vb{v}_0 = \vb{E}\times\vb{B}/E^2$, $\varphi = t - x$.
For the sake of simplicity we consider that the initial electron velocity lays in the $xy$-plane.
In that case the $y$-component of the electron velocity will always remain zero.
Let us write reduced motion equations~\eqref{eq.main1}--\eqref{eq.main2} in the considered configuration
\begin{align}
    & \dv{v}{t} =  \frac{E}{2\gamma}\left( v^2 - \frac{1}{\gamma^2} \right), \\
    & \dv{\gamma}{t} = -vE - F_{rr}(\chi),  \\
    & \dv{\varphi}{t} =\frac{1}{2}\left( v^2 + \frac{1}{\gamma^2} \right),
\end{align}
where $v=v_1$.
Changing integration variable from $t$ to $\varphi$ we get
\begin{align}
    \label{eq.pw.dvdf_init}
    \dv{v}{\varphi} = \frac{E}{2\gamma}\left( v^2 - \frac{1}{\gamma^2} \right) \frac{2}{v^2 + 1/\gamma^2} \\
    \label{eq.pw.dgdf_init}
    \dv{\gamma}{\varphi} = -(vE + F_{rr}(\chi)) \frac{2}{v^2 + 1/\gamma^2}
\end{align}
Expression for QED parameter $\chi$ is the following
\begin{equation}
    \chi = \frac{\gamma E}{E_S} \frac{1}{2} \left( v^2 + \frac{1}{\gamma^2} \right)
\end{equation}
As $\chi$ decreases due to radiation, eventually any electron will reach classical regime when $\chi \ll 1$.
In that case
\begin{equation}
    F_{rr}(\chi) = \frac{2}{3} \alpha E_S \chi^2 = A_{rr} \gamma^2 E^2 \left( v^2 + \frac{1}{\gamma^2} \right)^2
\end{equation}
Rewriting Eqs.~\eqref{eq.pw.dvdf_init}--\eqref{eq.pw.dgdf_init}
\begin{align}
     & \dv{v}{\varphi} = \frac{E}{\gamma}\frac{(v\gamma)^2 - 1}{(v\gamma)^2 + 1}                                        \\
     & \dv{\gamma}{\varphi} = -2\frac{(v\gamma)\gamma E}{(v\gamma)^2 + 1} - 2 A_{rr} E^2 \left( (v\gamma)^2 + 1 \right)
\end{align}
Introducing momentum $p = v\gamma$ we obtain final set of equations
\begin{align}
    \label{eq.pw.dgdf}
     & \dv{\gamma}{\varphi} = -2\frac{p\gamma E}{1 + p^2 } - 2 A_{rr} E^2 \left( 1 + p^2 \right) \\
    \label{eq.pw.dzdf}
     & \dv{p}{\varphi} = -E - 2 A_{rr} E^2 \frac{p}{\gamma}\left( 1 + p^2 \right)
\end{align}
Let us start solving Eqs.~\eqref{eq.pw.dgdf}--\eqref{eq.pw.dzdf} by assuming $A_{rr}=0$ and $E=E_0\cos\varphi$
\begin{align}
     & \dv{\gamma_{pw}}{\varphi} = -2\frac{p_{pw}\gamma_{pw} E_0\cos\varphi}{1 + {p_{pw}}^2 } \\
     & \dv{p_{pw}}{\varphi} = -E_0\cos\varphi
\end{align}
Eq.~\eqref{eq.pw.dzdf} has the following solution
\begin{equation}
    p_{pw} = -E_0\sin\varphi.
\end{equation}
Substituting this to Eq.~\eqref{eq.pw.dgdf} we get
\begin{align}
     & \dv{\gamma_{pw}}{\varphi} = \gamma_{pw}\frac{2 E_0^2\sin\varphi\cos\varphi}{1 + E_0^2\sin^2\varphi }  ,\\
     & \gamma_{pw} = \gamma_0 \left( 1 + E_0^2\sin^2\varphi \right) = \gamma_0 \left( 1 + {p_{pw}}^2 \right).
\end{align}
To find solution when $A\neq 0$ let us assume
\begin{align}
     & p = p_{pw} + u ,             \\
     & \gamma = \gamma_{pw} \Gamma.
\end{align}
where $u \ll p_{pw}$.
Eqs.~\eqref{eq.pw.dgdf}--\eqref{eq.pw.dzdf} transform to
\begin{align}
    \label{eq.pw.G_fin}
    &\begin{multlined}[c][7cm]
         \dv{\Gamma}{\varphi} = -2 \frac{A_{rr} E_0^2 \cos^2\varphi}{\gamma_0} - \\
         -2 u \Gamma E_0\cos\varphi \frac{ 1 - E_0^2 \cos^2\varphi}{\left( 1+E_0^2 \cos^2\varphi \right)^2} 
    \end{multlined},\\
     & \dv{u}{\varphi} = 2 A_{rr} E_0^3 \frac{\cos^2\varphi\sin\varphi}{\gamma_0\Gamma}.
\end{align}
Assuming that $\Gamma$ changes only slightly during single period it can be factored outside the integration, i.e.
\begin{equation}
    \begin{multlined}[c][7cm]
        u \approx \frac{2 A_{rr} E_0^3}{\gamma_0\Gamma} \int\limits_{0}^{\varphi}\cos^2\varphi'\sin\varphi'\dd\varphi' = \\
        =\frac{2 A_{rr} E_0^3}{3\gamma_0\Gamma}\left( 1 - \cos^3\varphi \right).
    \end{multlined}
\end{equation}
Substituting to Eq.~\eqref{eq.pw.G_fin}
\begin{equation}
    \begin{multlined}[c][7cm]
        \dv{\Gamma}{\varphi} = -2 \frac{A_{rr} E_0^2 \cos^2\varphi}{\gamma_0} - \frac{4 A_{rr} E_0^4}{3\gamma_0}\times \\
        \times \frac{ \cos\varphi\left( 1 - \cos^3\varphi \right)\left(1 - E_0^2\sin^2\varphi\right)}{\left( 1 + E_0^2\sin^2\varphi \right)^2}.
    \end{multlined}
\end{equation}
Since we assume $\Gamma$ changes on a time scale much larger than a single wave period we can substitute rhs of the above equation with its average value, i.e.
\begin{equation}
    \begin{multlined}[c][7cm]
         \dv{\Gamma}{\varphi} \approx \frac{1}{2\pi}\int\limits_{0}^{2\pi}\left[  -2 \frac{A_{rr} E_0^2 \cos^2\varphi}{\gamma_0} - \frac{4 A_{rr} E_0^4}{3\gamma_0} \times \right. \\
         \left. \times \frac{ \cos\varphi\left( 1 - \cos^3\varphi \right)\left(1 - E_0^2\sin^2\varphi\right)}{\left( 1 + E_0^2\sin^2\varphi \right)^2}  \right]\dd\varphi = \\
         = \frac{A_{rr}}{\gamma_0}\left( E_0^2 + 4 - 4\sqrt{1+E_0^2} \right) \stackrel{E_0 \gg 1}{\approx} \frac{A_{rr} E_0^2}{\gamma_0}.
    \end{multlined}
\end{equation}
So
\begin{align}
    \Gamma \approx 1 + \frac{A_{rr}E_0^2}{\gamma_0}\varphi.
\end{align}
And final solution
\begin{align}
     & \gamma \approx \gamma_0 \left( 1 + E_0^2\sin^2\varphi \right) \left( 1 + \frac{A_{rr}E_0^2}{\gamma_0}\varphi \right)  ,                  \\
     & v \approx - \frac{E_0\sin\varphi}{\gamma_0\left( 1 + E_0^2\sin^2\varphi \right) \left( 1 + \frac{A_{rr}E_0^2}{\gamma_0}\varphi \right)}.
\end{align}
To express the solution in terms of laboratory time instead of the wave phase we solve the following equation:
\begin{align}
     & 
     \begin{multlined}[c][7cm]
          \dv{\varphi}{t} = \frac{1}{2}\left( v^2 + \frac{1}{\gamma^2} \right) = \\
          =\frac{1}{2\gamma_0^2\left( 1 + E_0^2\sin^2\varphi \right) \left( 1 + \frac{A_{rr}E_0^2}{\gamma_0}\varphi \right)^2}
     \end{multlined}, \\
     & t = 2\gamma_0^2\int\left( 1 + E_0^2\sin^2\varphi \right) \left( 1 + \frac{A_{rr}E_0^2}{\gamma_0}\varphi \right)^2\dd\varphi.
\end{align}
Averaging over a wave period we get an approximate relation
\begin{align}
    \langle t \rangle \approx E_0^6 A_{rr}^2 \varphi^3.
\end{align}

\bibliography{main}

\begin{thebibliography}{85}%
\makeatletter
\providecommand \@ifxundefined [1]{%
 \@ifx{#1\undefined}
}%
\providecommand \@ifnum [1]{%
 \ifnum #1\expandafter \@firstoftwo
 \else \expandafter \@secondoftwo
 \fi
}%
\providecommand \@ifx [1]{%
 \ifx #1\expandafter \@firstoftwo
 \else \expandafter \@secondoftwo
 \fi
}%
\providecommand \natexlab [1]{#1}%
\providecommand \enquote  [1]{``#1''}%
\providecommand \bibnamefont  [1]{#1}%
\providecommand \bibfnamefont [1]{#1}%
\providecommand \citenamefont [1]{#1}%
\providecommand \href@noop [0]{\@secondoftwo}%
\providecommand \href [0]{\begingroup \@sanitize@url \@href}%
\providecommand \@href[1]{\@@startlink{#1}\@@href}%
\providecommand \@@href[1]{\endgroup#1\@@endlink}%
\providecommand \@sanitize@url [0]{\catcode `\\12\catcode `\$12\catcode
  `\&12\catcode `\#12\catcode `\^12\catcode `\_12\catcode `\%12\relax}%
\providecommand \@@startlink[1]{}%
\providecommand \@@endlink[0]{}%
\providecommand \url  [0]{\begingroup\@sanitize@url \@url }%
\providecommand \@url [1]{\endgroup\@href {#1}{\urlprefix }}%
\providecommand \urlprefix  [0]{URL }%
\providecommand \Eprint [0]{\href }%
\providecommand \doibase [0]{https://doi.org/}%
\providecommand \selectlanguage [0]{\@gobble}%
\providecommand \bibinfo  [0]{\@secondoftwo}%
\providecommand \bibfield  [0]{\@secondoftwo}%
\providecommand \translation [1]{[#1]}%
\providecommand \BibitemOpen [0]{}%
\providecommand \bibitemStop [0]{}%
\providecommand \bibitemNoStop [0]{.\EOS\space}%
\providecommand \EOS [0]{\spacefactor3000\relax}%
\providecommand \BibitemShut  [1]{\csname bibitem#1\endcsname}%
\let\auto@bib@innerbib\@empty
\bibitem [{ELI()}]{ELI}%
  \BibitemOpen
  \href {http://www.eli-laser.eu.} {\emph {\bibinfo {title} {The Extreme Light
  Infrastructure (ELI) official website: http://www.eli-laser.eu}}}\BibitemShut
  {NoStop}%
\bibitem [{\citenamefont {Gan}\ \emph {et~al.}(2021)\citenamefont {Gan},
  \citenamefont {Yu}, \citenamefont {Wang}, \citenamefont {Liu}, \citenamefont
  {Xu}, \citenamefont {Li}, \citenamefont {Li}, \citenamefont {Yu},
  \citenamefont {Wang}, \citenamefont {Liu} \emph {et~al.}}]{SULF}%
  \BibitemOpen
  \bibfield  {author} {\bibinfo {author} {\bibfnamefont {Z.}~\bibnamefont
  {Gan}}, \bibinfo {author} {\bibfnamefont {L.}~\bibnamefont {Yu}}, \bibinfo
  {author} {\bibfnamefont {C.}~\bibnamefont {Wang}}, \bibinfo {author}
  {\bibfnamefont {Y.}~\bibnamefont {Liu}}, \bibinfo {author} {\bibfnamefont
  {Y.}~\bibnamefont {Xu}}, \bibinfo {author} {\bibfnamefont {W.}~\bibnamefont
  {Li}}, \bibinfo {author} {\bibfnamefont {S.}~\bibnamefont {Li}}, \bibinfo
  {author} {\bibfnamefont {L.}~\bibnamefont {Yu}}, \bibinfo {author}
  {\bibfnamefont {X.}~\bibnamefont {Wang}}, \bibinfo {author} {\bibfnamefont
  {X.}~\bibnamefont {Liu}}, \emph {et~al.},\ }\bibfield  {title} {\enquote
  {\bibinfo {title} {The shanghai superintense ultrafast laser facility (sulf)
  project},}\ }in\ \href {https://doi.org/10.1007/978-3-030-75089-3_10} {\emph
  {\bibinfo {booktitle} {Progress in Ultrafast Intense Laser Science XVI}}}\
  (\bibinfo  {publisher} {Springer},\ \bibinfo {year} {2021})\ pp.\ \bibinfo
  {pages} {199--217}\BibitemShut {NoStop}%
\bibitem [{\citenamefont {Shao}\ \emph {et~al.}(2020)\citenamefont {Shao},
  \citenamefont {Li}, \citenamefont {Peng}, \citenamefont {Wang}, \citenamefont
  {Qian}, \citenamefont {Leng},\ and\ \citenamefont {Li}}]{SEL}%
  \BibitemOpen
  \bibfield  {author} {\bibinfo {author} {\bibfnamefont {B.}~\bibnamefont
  {Shao}}, \bibinfo {author} {\bibfnamefont {Y.}~\bibnamefont {Li}}, \bibinfo
  {author} {\bibfnamefont {Y.}~\bibnamefont {Peng}}, \bibinfo {author}
  {\bibfnamefont {P.}~\bibnamefont {Wang}}, \bibinfo {author} {\bibfnamefont
  {J.}~\bibnamefont {Qian}}, \bibinfo {author} {\bibfnamefont {Y.}~\bibnamefont
  {Leng}},\ and\ \bibinfo {author} {\bibfnamefont {R.}~\bibnamefont {Li}},\
  }\bibfield  {title} {\enquote {\bibinfo {title} {Broad-bandwidth
  high-temporal-contrast carrier-envelope-phase-stabilized laser seed for 100
  pw lasers},}\ }\href@noop {} {\bibfield  {journal} {\bibinfo  {journal}
  {Optics Letters}\ }\textbf {\bibinfo {volume} {45}},\ \bibinfo {pages}
  {2215--2218} (\bibinfo {year} {2020})}\BibitemShut {NoStop}%
\bibitem [{XCE()}]{XCELS}%
  \BibitemOpen
  \href {http://www.xcels.iapras.ru} {}\Eprint
  {https://arxiv.org/abs/{XCELS}~---~http://www.xcels.iapras.ru}
  {{XCELS}~---~http://www.xcels.iapras.ru} \BibitemShut {NoStop}%
\bibitem [{\citenamefont {Cole}\ \emph {et~al.}(2018)\citenamefont {Cole},
  \citenamefont {Behm}, \citenamefont {Gerstmayr}, \citenamefont {Blackburn},
  \citenamefont {Wood}, \citenamefont {Baird}, \citenamefont {Duff},
  \citenamefont {Harvey}, \citenamefont {Ilderton}, \citenamefont {Joglekar},
  \citenamefont {Krushelnick}, \citenamefont {Kuschel}, \citenamefont
  {Marklund}, \citenamefont {McKenna}, \citenamefont {Murphy}, \citenamefont
  {Poder}, \citenamefont {Ridgers}, \citenamefont {Samarin}, \citenamefont
  {Sarri}, \citenamefont {Symes}, \citenamefont {Thomas}, \citenamefont
  {Warwick}, \citenamefont {Zepf}, \citenamefont {Najmudin},\ and\
  \citenamefont {Mangles}}]{cole2018}%
  \BibitemOpen
  \bibfield  {author} {\bibinfo {author} {\bibfnamefont {J.~M.}\ \bibnamefont
  {Cole}}, \bibinfo {author} {\bibfnamefont {K.~T.}\ \bibnamefont {Behm}},
  \bibinfo {author} {\bibfnamefont {E.}~\bibnamefont {Gerstmayr}}, \bibinfo
  {author} {\bibfnamefont {T.~G.}\ \bibnamefont {Blackburn}}, \bibinfo {author}
  {\bibfnamefont {J.~C.}\ \bibnamefont {Wood}}, \bibinfo {author}
  {\bibfnamefont {C.~D.}\ \bibnamefont {Baird}}, \bibinfo {author}
  {\bibfnamefont {M.~J.}\ \bibnamefont {Duff}}, \bibinfo {author}
  {\bibfnamefont {C.}~\bibnamefont {Harvey}}, \bibinfo {author} {\bibfnamefont
  {A.}~\bibnamefont {Ilderton}}, \bibinfo {author} {\bibfnamefont {A.~S.}\
  \bibnamefont {Joglekar}}, \bibinfo {author} {\bibfnamefont {K.}~\bibnamefont
  {Krushelnick}}, \bibinfo {author} {\bibfnamefont {S.}~\bibnamefont
  {Kuschel}}, \bibinfo {author} {\bibfnamefont {M.}~\bibnamefont {Marklund}},
  \bibinfo {author} {\bibfnamefont {P.}~\bibnamefont {McKenna}}, \bibinfo
  {author} {\bibfnamefont {C.~D.}\ \bibnamefont {Murphy}}, \bibinfo {author}
  {\bibfnamefont {K.}~\bibnamefont {Poder}}, \bibinfo {author} {\bibfnamefont
  {C.~P.}\ \bibnamefont {Ridgers}}, \bibinfo {author} {\bibfnamefont {G.~M.}\
  \bibnamefont {Samarin}}, \bibinfo {author} {\bibfnamefont {G.}~\bibnamefont
  {Sarri}}, \bibinfo {author} {\bibfnamefont {D.~R.}\ \bibnamefont {Symes}},
  \bibinfo {author} {\bibfnamefont {A.~G.~R.}\ \bibnamefont {Thomas}}, \bibinfo
  {author} {\bibfnamefont {J.}~\bibnamefont {Warwick}}, \bibinfo {author}
  {\bibfnamefont {M.}~\bibnamefont {Zepf}}, \bibinfo {author} {\bibfnamefont
  {Z.}~\bibnamefont {Najmudin}},\ and\ \bibinfo {author} {\bibfnamefont
  {S.~P.~D.}\ \bibnamefont {Mangles}},\ }\bibfield  {title} {\enquote {\bibinfo
  {title} {Experimental evidence of radiation reaction in the collision of a
  high-intensity laser pulse with a laser-wakefield accelerated electron
  beam},}\ }\href {https://doi.org/10.1103/PhysRevX.8.011020} {\bibfield
  {journal} {\bibinfo  {journal} {Phys. Rev. X}\ }\textbf {\bibinfo {volume}
  {8}},\ \bibinfo {pages} {011020} (\bibinfo {year} {2018})}\BibitemShut
  {NoStop}%
\bibitem [{\citenamefont {Poder}\ \emph {et~al.}(2018)\citenamefont {Poder},
  \citenamefont {Tamburini}, \citenamefont {Sarri}, \citenamefont {Di~Piazza},
  \citenamefont {Kuschel}, \citenamefont {Baird}, \citenamefont {Behm},
  \citenamefont {Bohlen}, \citenamefont {Cole}, \citenamefont {Corvan},
  \citenamefont {Duff}, \citenamefont {Gerstmayr}, \citenamefont {Keitel},
  \citenamefont {Krushelnick}, \citenamefont {Mangles}, \citenamefont
  {McKenna}, \citenamefont {Murphy}, \citenamefont {Najmudin}, \citenamefont
  {Ridgers}, \citenamefont {Samarin}, \citenamefont {Symes}, \citenamefont
  {Thomas}, \citenamefont {Warwick},\ and\ \citenamefont {Zepf}}]{poder2018}%
  \BibitemOpen
  \bibfield  {author} {\bibinfo {author} {\bibfnamefont {K.}~\bibnamefont
  {Poder}}, \bibinfo {author} {\bibfnamefont {M.}~\bibnamefont {Tamburini}},
  \bibinfo {author} {\bibfnamefont {G.}~\bibnamefont {Sarri}}, \bibinfo
  {author} {\bibfnamefont {A.}~\bibnamefont {Di~Piazza}}, \bibinfo {author}
  {\bibfnamefont {S.}~\bibnamefont {Kuschel}}, \bibinfo {author} {\bibfnamefont
  {C.~D.}\ \bibnamefont {Baird}}, \bibinfo {author} {\bibfnamefont
  {K.}~\bibnamefont {Behm}}, \bibinfo {author} {\bibfnamefont {S.}~\bibnamefont
  {Bohlen}}, \bibinfo {author} {\bibfnamefont {J.~M.}\ \bibnamefont {Cole}},
  \bibinfo {author} {\bibfnamefont {D.~J.}\ \bibnamefont {Corvan}}, \bibinfo
  {author} {\bibfnamefont {M.}~\bibnamefont {Duff}}, \bibinfo {author}
  {\bibfnamefont {E.}~\bibnamefont {Gerstmayr}}, \bibinfo {author}
  {\bibfnamefont {C.~H.}\ \bibnamefont {Keitel}}, \bibinfo {author}
  {\bibfnamefont {K.}~\bibnamefont {Krushelnick}}, \bibinfo {author}
  {\bibfnamefont {S.~P.~D.}\ \bibnamefont {Mangles}}, \bibinfo {author}
  {\bibfnamefont {P.}~\bibnamefont {McKenna}}, \bibinfo {author} {\bibfnamefont
  {C.~D.}\ \bibnamefont {Murphy}}, \bibinfo {author} {\bibfnamefont
  {Z.}~\bibnamefont {Najmudin}}, \bibinfo {author} {\bibfnamefont {C.~P.}\
  \bibnamefont {Ridgers}}, \bibinfo {author} {\bibfnamefont {G.~M.}\
  \bibnamefont {Samarin}}, \bibinfo {author} {\bibfnamefont {D.~R.}\
  \bibnamefont {Symes}}, \bibinfo {author} {\bibfnamefont {A.~G.~R.}\
  \bibnamefont {Thomas}}, \bibinfo {author} {\bibfnamefont {J.}~\bibnamefont
  {Warwick}},\ and\ \bibinfo {author} {\bibfnamefont {M.}~\bibnamefont
  {Zepf}},\ }\bibfield  {title} {\enquote {\bibinfo {title} {Experimental
  signatures of the quantum nature of radiation reaction in the field of an
  ultraintense laser},}\ }\href {https://doi.org/10.1103/PhysRevX.8.031004}
  {\bibfield  {journal} {\bibinfo  {journal} {Phys. Rev. X}\ }\textbf {\bibinfo
  {volume} {8}},\ \bibinfo {pages} {031004} (\bibinfo {year}
  {2018})}\BibitemShut {NoStop}%
\bibitem [{\citenamefont {Tamburini}\ \emph {et~al.}(2010)\citenamefont
  {Tamburini}, \citenamefont {Pegoraro}, \citenamefont {Piazza}, \citenamefont
  {Keitel},\ and\ \citenamefont {Macchi}}]{Tamburini10}%
  \BibitemOpen
  \bibfield  {author} {\bibinfo {author} {\bibfnamefont {M.}~\bibnamefont
  {Tamburini}}, \bibinfo {author} {\bibfnamefont {F.}~\bibnamefont {Pegoraro}},
  \bibinfo {author} {\bibfnamefont {A.~D.}\ \bibnamefont {Piazza}}, \bibinfo
  {author} {\bibfnamefont {C.~H.}\ \bibnamefont {Keitel}},\ and\ \bibinfo
  {author} {\bibfnamefont {A.}~\bibnamefont {Macchi}},\ }\bibfield  {title}
  {\enquote {\bibinfo {title} {Radiation reaction effects on radiation pressure
  acceleration},}\ }\href {https://doi.org/10.1088/1367-2630/12/12/123005}
  {\bibfield  {journal} {\bibinfo  {journal} {New Journal of Physics}\ }\textbf
  {\bibinfo {volume} {12}},\ \bibinfo {pages} {123005} (\bibinfo {year}
  {2010})}\BibitemShut {NoStop}%
\bibitem [{\citenamefont {Tamburini}\ \emph {et~al.}(2012)\citenamefont
  {Tamburini}, \citenamefont {Liseykina}, \citenamefont {Pegoraro},\ and\
  \citenamefont {Macchi}}]{Tamburini12}%
  \BibitemOpen
  \bibfield  {author} {\bibinfo {author} {\bibfnamefont {M.}~\bibnamefont
  {Tamburini}}, \bibinfo {author} {\bibfnamefont {T.~V.}\ \bibnamefont
  {Liseykina}}, \bibinfo {author} {\bibfnamefont {F.}~\bibnamefont
  {Pegoraro}},\ and\ \bibinfo {author} {\bibfnamefont {A.}~\bibnamefont
  {Macchi}},\ }\bibfield  {title} {\enquote {\bibinfo {title}
  {Radiation-pressure-dominant acceleration: Polarization and radiation
  reaction effects and energy increase in three-dimensional simulations},}\
  }\href {https://doi.org/10.1103/PhysRevE.85.016407} {\bibfield  {journal}
  {\bibinfo  {journal} {Physical Review E}\ }\textbf {\bibinfo {volume} {85}},\
  \bibinfo {pages} {016407} (\bibinfo {year} {2012})}\BibitemShut {NoStop}%
\bibitem [{\citenamefont {Kostyukov}, \citenamefont {Nerush},\ and\
  \citenamefont {Litvak}(2012)}]{kostyukov2012radiative}%
  \BibitemOpen
  \bibfield  {author} {\bibinfo {author} {\bibfnamefont {I.~{\relax Yu}.}\
  \bibnamefont {Kostyukov}}, \bibinfo {author} {\bibfnamefont {E.~N.}\
  \bibnamefont {Nerush}},\ and\ \bibinfo {author} {\bibfnamefont {A.~G.}\
  \bibnamefont {Litvak}},\ }\bibfield  {title} {\enquote {\bibinfo {title}
  {Radiative damping in plasma-based accelerators},}\ }\href
  {https://doi.org/10.1103/PhysRevSTAB.15.111001} {\bibfield  {journal}
  {\bibinfo  {journal} {Physical Review Special Topics-Accelerators and Beams}\
  }\textbf {\bibinfo {volume} {15}},\ \bibinfo {pages} {111001} (\bibinfo
  {year} {2012})}\BibitemShut {NoStop}%
\bibitem [{\citenamefont {Capdessus}, \citenamefont {d'Humi{\`{e}}res},\ and\
  \citenamefont {Tikhonchuk}(2012)}]{Capdessus12}%
  \BibitemOpen
  \bibfield  {author} {\bibinfo {author} {\bibfnamefont {R.}~\bibnamefont
  {Capdessus}}, \bibinfo {author} {\bibfnamefont {E.}~\bibnamefont
  {d'Humi{\`{e}}res}},\ and\ \bibinfo {author} {\bibfnamefont {V.~T.}\
  \bibnamefont {Tikhonchuk}},\ }\bibfield  {title} {\enquote {\bibinfo {title}
  {Modeling of radiation losses in ultrahigh power laser-matter interaction},}\
  }\href {https://doi.org/10.1103/PhysRevE.86.036401} {\bibfield  {journal}
  {\bibinfo  {journal} {Physical Review E}\ }\textbf {\bibinfo {volume} {86}},\
  \bibinfo {pages} {036401} (\bibinfo {year} {2012})}\BibitemShut {NoStop}%
\bibitem [{\citenamefont {Capdessus}\ and\ \citenamefont
  {McKenna}(2015)}]{Capdessus15}%
  \BibitemOpen
  \bibfield  {author} {\bibinfo {author} {\bibfnamefont {R.}~\bibnamefont
  {Capdessus}}\ and\ \bibinfo {author} {\bibfnamefont {P.}~\bibnamefont
  {McKenna}},\ }\bibfield  {title} {\enquote {\bibinfo {title} {Influence of
  radiation reaction force on ultraintense laser-driven ion acceleration},}\
  }\href {https://doi.org/10.1103/PhysRevE.91.053105} {\bibfield  {journal}
  {\bibinfo  {journal} {Physical Review E}\ }\textbf {\bibinfo {volume} {91}},\
  \bibinfo {pages} {053105} (\bibinfo {year} {2015})}\BibitemShut {NoStop}%
\bibitem [{\citenamefont {Nerush}\ and\ \citenamefont
  {Kostyukov}(2015)}]{Nerush15}%
  \BibitemOpen
  \bibfield  {author} {\bibinfo {author} {\bibfnamefont {E.~N.}\ \bibnamefont
  {Nerush}}\ and\ \bibinfo {author} {\bibfnamefont {I.~Y.}\ \bibnamefont
  {Kostyukov}},\ }\bibfield  {title} {\enquote {\bibinfo {title} {Laser-driven
  hole boring and gamma-ray emission in high-density plasmas},}\ }\href
  {https://doi.org/10.1088/0741-3335/57/3/035007} {\bibfield  {journal}
  {\bibinfo  {journal} {Plasma Physics and Controlled Fusion}\ }\textbf
  {\bibinfo {volume} {57}},\ \bibinfo {pages} {035007} (\bibinfo {year}
  {2015})}\BibitemShut {NoStop}%
\bibitem [{\citenamefont {Gelfer}, \citenamefont {Fedotov},\ and\ \citenamefont
  {Weber}(2018)}]{Gelfer18a}%
  \BibitemOpen
  \bibfield  {author} {\bibinfo {author} {\bibfnamefont {E.~G.}\ \bibnamefont
  {Gelfer}}, \bibinfo {author} {\bibfnamefont {A.~M.}\ \bibnamefont
  {Fedotov}},\ and\ \bibinfo {author} {\bibfnamefont {S.}~\bibnamefont
  {Weber}},\ }\bibfield  {title} {\enquote {\bibinfo {title} {Theory and
  simulations of radiation friction induced enhancement of laser-driven
  longitudinal fields},}\ }\href {https://doi.org/10.1088/1361-6587/aabb12}
  {\bibfield  {journal} {\bibinfo  {journal} {Plasma Physics and Controlled
  Fusion}\ }\textbf {\bibinfo {volume} {60}},\ \bibinfo {pages} {064005}
  (\bibinfo {year} {2018})}\BibitemShut {NoStop}%
\bibitem [{\citenamefont {Gelfer}, \citenamefont {Elkina},\ and\ \citenamefont
  {Fedotov}(2018)}]{Gelfer18b}%
  \BibitemOpen
  \bibfield  {author} {\bibinfo {author} {\bibfnamefont {E.}~\bibnamefont
  {Gelfer}}, \bibinfo {author} {\bibfnamefont {N.}~\bibnamefont {Elkina}},\
  and\ \bibinfo {author} {\bibfnamefont {A.}~\bibnamefont {Fedotov}},\
  }\bibfield  {title} {\enquote {\bibinfo {title} {Unexpected impact of
  radiation friction: enhancing production of longitudinal plasma waves},}\
  }\href {https://doi.org/10.1038/s41598-018-24930-x} {\bibfield  {journal}
  {\bibinfo  {journal} {Scientific Reports}\ }\textbf {\bibinfo {volume} {8}},\
  \bibinfo {pages} {6478} (\bibinfo {year} {2018})}\BibitemShut {NoStop}%
\bibitem [{\citenamefont {{G}elfer}, \citenamefont {{F}edotov},\ and\
  \citenamefont {{W}eber}(2021)}]{gelfer2021ions}%
  \BibitemOpen
  \bibfield  {author} {\bibinfo {author} {\bibfnamefont {E.~G.}\ \bibnamefont
  {{G}elfer}}, \bibinfo {author} {\bibfnamefont {A.~M.}\ \bibnamefont
  {{F}edotov}},\ and\ \bibinfo {author} {\bibfnamefont {S.}~\bibnamefont
  {{W}eber}},\ }\bibfield  {title} {\enquote {\bibinfo {title} {{R}adiation
  induced acceleration of ions in a laser irradiated transparent foil},}\
  }\href {https://doi.org/10.1088/1367-2630/ac1a97} {\bibfield  {journal}
  {\bibinfo  {journal} {{N}ew {J}ournal of {P}hysics}\ }\textbf {\bibinfo
  {volume} {23}},\ \bibinfo {pages} {095002} (\bibinfo {year}
  {2021})}\BibitemShut {NoStop}%
\bibitem [{\citenamefont {Golovanov}, \citenamefont {Nerush},\ and\
  \citenamefont {Kostyukov}(2022)}]{golovanov2021radiation}%
  \BibitemOpen
  \bibfield  {author} {\bibinfo {author} {\bibfnamefont {A.~A.}\ \bibnamefont
  {Golovanov}}, \bibinfo {author} {\bibfnamefont {E.~N.}\ \bibnamefont
  {Nerush}},\ and\ \bibinfo {author} {\bibfnamefont {I.~{\relax Yu}.}\
  \bibnamefont {Kostyukov}},\ }\bibfield  {title} {\enquote {\bibinfo {title}
  {Radiation reaction-dominated regime of wakefield acceleration},}\ }\href
  {https://doi.org/10.1088/1367-2630/ac53b9} {\bibfield  {journal} {\bibinfo
  {journal} {New Journal of Physics}\ }\textbf {\bibinfo {volume} {24}},\
  \bibinfo {pages} {033011} (\bibinfo {year} {2022})}\BibitemShut {NoStop}%
\bibitem [{\citenamefont {Grismayer}\ \emph {et~al.}(2016)\citenamefont
  {Grismayer}, \citenamefont {Vranic}, \citenamefont {Martins}, \citenamefont
  {Fonseca},\ and\ \citenamefont {Silva}}]{Grismayer16}%
  \BibitemOpen
  \bibfield  {author} {\bibinfo {author} {\bibfnamefont {T.}~\bibnamefont
  {Grismayer}}, \bibinfo {author} {\bibfnamefont {M.}~\bibnamefont {Vranic}},
  \bibinfo {author} {\bibfnamefont {J.~L.}\ \bibnamefont {Martins}}, \bibinfo
  {author} {\bibfnamefont {R.~A.}\ \bibnamefont {Fonseca}},\ and\ \bibinfo
  {author} {\bibfnamefont {L.~O.}\ \bibnamefont {Silva}},\ }\bibfield  {title}
  {\enquote {\bibinfo {title} {Laser absorption via quantum electrodynamics
  cascades in counter propagating laser pulses},}\ }\href
  {https://doi.org/10.1063/1.4950841} {\bibfield  {journal} {\bibinfo
  {journal} {Physics of Plasmas}\ }\textbf {\bibinfo {volume} {23}},\ \bibinfo
  {pages} {056706} (\bibinfo {year} {2016})}\BibitemShut {NoStop}%
\bibitem [{\citenamefont {Zhang}, \citenamefont {Ridgers},\ and\ \citenamefont
  {Thomas}(2015)}]{Zhang15}%
  \BibitemOpen
  \bibfield  {author} {\bibinfo {author} {\bibfnamefont {P.}~\bibnamefont
  {Zhang}}, \bibinfo {author} {\bibfnamefont {C.~P.}\ \bibnamefont {Ridgers}},\
  and\ \bibinfo {author} {\bibfnamefont {A.~G.~R.}\ \bibnamefont {Thomas}},\
  }\bibfield  {title} {\enquote {\bibinfo {title} {The effect of nonlinear
  quantum electrodynamics on relativistic transparency and laser absorption in
  ultra-relativistic plasmas},}\ }\href
  {https://doi.org/10.1088/1367-2630/17/4/043051} {\bibfield  {journal}
  {\bibinfo  {journal} {New Journal of Physics}\ }\textbf {\bibinfo {volume}
  {17}},\ \bibinfo {pages} {043051} (\bibinfo {year} {2015})}\BibitemShut
  {NoStop}%
\bibitem [{\citenamefont {Liseykina}, \citenamefont {Popruzhenko},\ and\
  \citenamefont {Macchi}(2016)}]{Liseykina16}%
  \BibitemOpen
  \bibfield  {author} {\bibinfo {author} {\bibfnamefont {T.~V.}\ \bibnamefont
  {Liseykina}}, \bibinfo {author} {\bibfnamefont {S.~V.}\ \bibnamefont
  {Popruzhenko}},\ and\ \bibinfo {author} {\bibfnamefont {A.}~\bibnamefont
  {Macchi}},\ }\bibfield  {title} {\enquote {\bibinfo {title} {Inverse faraday
  effect driven by radiation friction},}\ }\href
  {https://doi.org/10.1088/1367-2630/18/7/072001} {\bibfield  {journal}
  {\bibinfo  {journal} {New Journal of Physics}\ }\textbf {\bibinfo {volume}
  {18}},\ \bibinfo {pages} {072001} (\bibinfo {year} {2016})}\BibitemShut
  {NoStop}%
\bibitem [{\citenamefont {{L}iseykina}, \citenamefont {{M}acchi},\ and\
  \citenamefont {{P}opruzhenko}(2021)}]{liseykina2021IFE}%
  \BibitemOpen
  \bibfield  {author} {\bibinfo {author} {\bibfnamefont {T.~V.}\ \bibnamefont
  {{L}iseykina}}, \bibinfo {author} {\bibfnamefont {A.}~\bibnamefont
  {{M}acchi}},\ and\ \bibinfo {author} {\bibfnamefont {S.~V.}\ \bibnamefont
  {{P}opruzhenko}},\ }\bibfield  {title} {\enquote {\bibinfo {title} {{Q}uantum
  effects on radiation friction driven magnetic field generation},}\ }\href
  {https://doi.org/10.1140/epjp/s13360-020-01030-2} {\bibfield  {journal}
  {\bibinfo  {journal} {{T}he {E}uropean {P}hysical {J}ournal {P}lus}\ }\textbf
  {\bibinfo {volume} {136}},\ \bibinfo {pages} {170} (\bibinfo {year}
  {2021})}\BibitemShut {NoStop}%
\bibitem [{\citenamefont {{S}amsonov}, \citenamefont {{N}erush},\ and\
  \citenamefont {{K}ostyukov}(2021)}]{Samsonov2021IFE}%
  \BibitemOpen
  \bibfield  {author} {\bibinfo {author} {\bibfnamefont {A.~S.}\ \bibnamefont
  {{S}amsonov}}, \bibinfo {author} {\bibfnamefont {E.~N.}\ \bibnamefont
  {{N}erush}},\ and\ \bibinfo {author} {\bibfnamefont {I.~{\relax Yu}.}\
  \bibnamefont {{K}ostyukov}},\ }\bibfield  {title} {\enquote {\bibinfo {title}
  {{E}ffect of {e}lectron–{p}ositron plasma production on the generation of a
  magnetic field in laser-plasma interactions},}\ }\href
  {https://doi.org/10.1070/qel17601} {\bibfield  {journal} {\bibinfo  {journal}
  {{Q}uantum {E}lectronics}\ }\textbf {\bibinfo {volume} {51}},\ \bibinfo
  {pages} {861--865} (\bibinfo {year} {2021})}\BibitemShut {NoStop}%
\bibitem [{\citenamefont {{D}el {S}orbo}\ \emph {et~al.}(2017)\citenamefont
  {{D}el {S}orbo}, \citenamefont {{S}eipt}, \citenamefont {{B}lackburn},
  \citenamefont {{T}homas}, \citenamefont {{M}urphy}, \citenamefont {{K}irk},\
  and\ \citenamefont {{R}idgers}}]{DelSorbo2017Spin}%
  \BibitemOpen
  \bibfield  {author} {\bibinfo {author} {\bibfnamefont {D.}~\bibnamefont
  {{D}el {S}orbo}}, \bibinfo {author} {\bibfnamefont {D.}~\bibnamefont
  {{S}eipt}}, \bibinfo {author} {\bibfnamefont {T.~G.}\ \bibnamefont
  {{B}lackburn}}, \bibinfo {author} {\bibfnamefont {A.~G.~R.}\ \bibnamefont
  {{T}homas}}, \bibinfo {author} {\bibfnamefont {C.~D.}\ \bibnamefont
  {{M}urphy}}, \bibinfo {author} {\bibfnamefont {J.~G.}\ \bibnamefont
  {{K}irk}},\ and\ \bibinfo {author} {\bibfnamefont {C.~P.}\ \bibnamefont
  {{R}idgers}},\ }\bibfield  {title} {\enquote {\bibinfo {title} {{S}pin
  polarization of electrons by ultraintense lasers},}\ }\href
  {https://doi.org/10.1103/physreva.96.043407} {\bibfield  {journal} {\bibinfo
  {journal} {{P}hysical {R}eview {A}}\ }\textbf {\bibinfo {volume} {96}},\
  \bibinfo {pages} {043407} (\bibinfo {year} {2017})}\BibitemShut {NoStop}%
\bibitem [{\citenamefont {{D}el {S}orbo}\ \emph {et~al.}(2018)\citenamefont
  {{D}el {S}orbo}, \citenamefont {{S}eipt}, \citenamefont {{T}homas},\ and\
  \citenamefont {{R}idgers}}]{DelSorbo2018spin}%
  \BibitemOpen
  \bibfield  {author} {\bibinfo {author} {\bibfnamefont {D.}~\bibnamefont
  {{D}el {S}orbo}}, \bibinfo {author} {\bibfnamefont {D.}~\bibnamefont
  {{S}eipt}}, \bibinfo {author} {\bibfnamefont {A.~G.~R.}\ \bibnamefont
  {{T}homas}},\ and\ \bibinfo {author} {\bibfnamefont {C.~P.}\ \bibnamefont
  {{R}idgers}},\ }\bibfield  {title} {\enquote {\bibinfo {title} {{E}lectron
  spin polarization in realistic trajectories around the magnetic node of two
  counter-propagating, circularly polarized, ultra-intense lasers},}\ }\href
  {https://doi.org/10.1088/1361-6587/aab979} {\bibfield  {journal} {\bibinfo
  {journal} {{P}lasma {P}hysics and {C}ontrolled {F}usion}\ }\textbf {\bibinfo
  {volume} {60}},\ \bibinfo {pages} {064003} (\bibinfo {year}
  {2018})}\BibitemShut {NoStop}%
\bibitem [{\citenamefont {{C}hen}\ \emph {et~al.}(2019)\citenamefont {{C}hen},
  \citenamefont {{H}e}, \citenamefont {{S}haisultanov}, \citenamefont
  {{H}atsagortsyan},\ and\ \citenamefont {{K}eitel}}]{Chen2019spin}%
  \BibitemOpen
  \bibfield  {author} {\bibinfo {author} {\bibfnamefont {Y.-Y.}\ \bibnamefont
  {{C}hen}}, \bibinfo {author} {\bibfnamefont {P.-L.}\ \bibnamefont {{H}e}},
  \bibinfo {author} {\bibfnamefont {R.}~\bibnamefont {{S}haisultanov}},
  \bibinfo {author} {\bibfnamefont {K.~Z.}\ \bibnamefont {{H}atsagortsyan}},\
  and\ \bibinfo {author} {\bibfnamefont {C.~H.}\ \bibnamefont {{K}eitel}},\
  }\bibfield  {title} {\enquote {\bibinfo {title} {{P}olarized {P}ositron
  {B}eams via {I}ntense {T}wo-{C}olor {L}aser {P}ulses},}\ }\href
  {https://doi.org/10.1103/physrevlett.123.174801} {\bibfield  {journal}
  {\bibinfo  {journal} {{P}hysical {R}eview {L}etters}\ }\textbf {\bibinfo
  {volume} {123}},\ \bibinfo {pages} {174801} (\bibinfo {year}
  {2019})}\BibitemShut {NoStop}%
\bibitem [{\citenamefont {Seipt}\ \emph {et~al.}(2019)\citenamefont {Seipt},
  \citenamefont {Del~Sorbo}, \citenamefont {Ridgers},\ and\ \citenamefont
  {Thomas}}]{Seipt2019spin}%
  \BibitemOpen
  \bibfield  {author} {\bibinfo {author} {\bibfnamefont {D.}~\bibnamefont
  {Seipt}}, \bibinfo {author} {\bibfnamefont {D.}~\bibnamefont {Del~Sorbo}},
  \bibinfo {author} {\bibfnamefont {C.~P.}\ \bibnamefont {Ridgers}},\ and\
  \bibinfo {author} {\bibfnamefont {A.~G.~R.}\ \bibnamefont {Thomas}},\
  }\bibfield  {title} {\enquote {\bibinfo {title} {Ultrafast polarization of an
  electron beam in an intense bichromatic laser field},}\ }\href
  {https://doi.org/10.1103/PhysRevA.100.061402} {\bibfield  {journal} {\bibinfo
   {journal} {Physical Review A}\ }\textbf {\bibinfo {volume} {100}},\ \bibinfo
  {pages} {061402} (\bibinfo {year} {2019})}\BibitemShut {NoStop}%
\bibitem [{\citenamefont {{W}u}\ \emph {et~al.}(2019)\citenamefont {{W}u},
  \citenamefont {{J}i}, \citenamefont {{G}eng}, \citenamefont {{Y}u},
  \citenamefont {{W}ang}, \citenamefont {{F}eng}, \citenamefont {{G}uo},
  \citenamefont {{W}ang}, \citenamefont {{Q}in}, \citenamefont {{Y}an},
  \citenamefont {{Z}hang}, \citenamefont {{T}homas}, \citenamefont {{H}ützen},
  \citenamefont {{B}üscher}, \citenamefont {{R}akitzis}, \citenamefont
  {{P}ukhov}, \citenamefont {{S}hen},\ and\ \citenamefont {{L}i}}]{Wu2019spin}%
  \BibitemOpen
  \bibfield  {author} {\bibinfo {author} {\bibfnamefont {Y.}~\bibnamefont
  {{W}u}}, \bibinfo {author} {\bibfnamefont {L.}~\bibnamefont {{J}i}}, \bibinfo
  {author} {\bibfnamefont {X.}~\bibnamefont {{G}eng}}, \bibinfo {author}
  {\bibfnamefont {Q.}~\bibnamefont {{Y}u}}, \bibinfo {author} {\bibfnamefont
  {N.}~\bibnamefont {{W}ang}}, \bibinfo {author} {\bibfnamefont
  {B.}~\bibnamefont {{F}eng}}, \bibinfo {author} {\bibfnamefont
  {Z.}~\bibnamefont {{G}uo}}, \bibinfo {author} {\bibfnamefont
  {W.}~\bibnamefont {{W}ang}}, \bibinfo {author} {\bibfnamefont
  {C.}~\bibnamefont {{Q}in}}, \bibinfo {author} {\bibfnamefont
  {X.}~\bibnamefont {{Y}an}}, \bibinfo {author} {\bibfnamefont
  {L.}~\bibnamefont {{Z}hang}}, \bibinfo {author} {\bibfnamefont
  {J.}~\bibnamefont {{T}homas}}, \bibinfo {author} {\bibfnamefont
  {A.}~\bibnamefont {{H}ützen}}, \bibinfo {author} {\bibfnamefont
  {M.}~\bibnamefont {{B}üscher}}, \bibinfo {author} {\bibfnamefont {T.~P.}\
  \bibnamefont {{R}akitzis}}, \bibinfo {author} {\bibfnamefont
  {A.}~\bibnamefont {{P}ukhov}}, \bibinfo {author} {\bibfnamefont
  {B.}~\bibnamefont {{S}hen}},\ and\ \bibinfo {author} {\bibfnamefont
  {R.}~\bibnamefont {{L}i}},\ }\bibfield  {title} {\enquote {\bibinfo {title}
  {{P}olarized electron-beam acceleration driven by vortex laser pulses},}\
  }\href {https://doi.org/10.1088/1367-2630/ab2fd7} {\bibfield  {journal}
  {\bibinfo  {journal} {{N}ew {J}ournal of {P}hysics}\ }\textbf {\bibinfo
  {volume} {21}},\ \bibinfo {pages} {073052} (\bibinfo {year}
  {2019})}\BibitemShut {NoStop}%
\bibitem [{\citenamefont {{L}i}\ \emph {et~al.}(2019)\citenamefont {{L}i},
  \citenamefont {{S}haisultanov}, \citenamefont {{H}atsagortsyan},
  \citenamefont {{W}an}, \citenamefont {{K}eitel},\ and\ \citenamefont
  {{L}i}}]{Li2019spin}%
  \BibitemOpen
  \bibfield  {author} {\bibinfo {author} {\bibfnamefont {Y.-F.}\ \bibnamefont
  {{L}i}}, \bibinfo {author} {\bibfnamefont {R.}~\bibnamefont
  {{S}haisultanov}}, \bibinfo {author} {\bibfnamefont {K.~Z.}\ \bibnamefont
  {{H}atsagortsyan}}, \bibinfo {author} {\bibfnamefont {F.}~\bibnamefont
  {{W}an}}, \bibinfo {author} {\bibfnamefont {C.~H.}\ \bibnamefont
  {{K}eitel}},\ and\ \bibinfo {author} {\bibfnamefont {J.-X.}\ \bibnamefont
  {{L}i}},\ }\bibfield  {title} {\enquote {\bibinfo {title}
  {{U}ltrarelativistic {E}lectron-{B}eam {P}olarization in {S}ingle-{S}hot
  {I}nteraction with an {U}ltraintense {L}aser {P}ulse},}\ }\href
  {https://doi.org/10.1103/physrevlett.122.154801} {\bibfield  {journal}
  {\bibinfo  {journal} {{P}hysical {R}eview {L}etters}\ }\textbf {\bibinfo
  {volume} {122}},\ \bibinfo {pages} {154801} (\bibinfo {year}
  {2019})}\BibitemShut {NoStop}%
\bibitem [{\citenamefont {{L}i}\ \emph {et~al.}(2020)\citenamefont {{L}i},
  \citenamefont {{C}hen}, \citenamefont {{W}ang},\ and\ \citenamefont
  {{H}u}}]{Li2020spin}%
  \BibitemOpen
  \bibfield  {author} {\bibinfo {author} {\bibfnamefont {Y.-F.}\ \bibnamefont
  {{L}i}}, \bibinfo {author} {\bibfnamefont {Y.-Y.}\ \bibnamefont {{C}hen}},
  \bibinfo {author} {\bibfnamefont {W.-M.}\ \bibnamefont {{W}ang}},\ and\
  \bibinfo {author} {\bibfnamefont {H.-S.}\ \bibnamefont {{H}u}},\ }\bibfield
  {title} {\enquote {\bibinfo {title} {{P}roduction of {H}ighly {P}olarized
  {P}ositron {B}eams via {H}elicity {T}ransfer from {P}olarized {E}lectrons in
  a {S}trong {L}aser {F}ield},}\ }\href
  {https://doi.org/10.1103/physrevlett.125.044802} {\bibfield  {journal}
  {\bibinfo  {journal} {{P}hysical {R}eview {L}etters}\ }\textbf {\bibinfo
  {volume} {125}},\ \bibinfo {pages} {044802} (\bibinfo {year}
  {2020})}\BibitemShut {NoStop}%
\bibitem [{\citenamefont {{W}an}\ \emph {et~al.}(2020)\citenamefont {{W}an},
  \citenamefont {{S}haisultanov}, \citenamefont {{L}i}, \citenamefont
  {{H}atsagortsyan}, \citenamefont {{K}eitel},\ and\ \citenamefont
  {{L}i}}]{Wan2020spin}%
  \BibitemOpen
  \bibfield  {author} {\bibinfo {author} {\bibfnamefont {F.}~\bibnamefont
  {{W}an}}, \bibinfo {author} {\bibfnamefont {R.}~\bibnamefont
  {{S}haisultanov}}, \bibinfo {author} {\bibfnamefont {Y.-F.}\ \bibnamefont
  {{L}i}}, \bibinfo {author} {\bibfnamefont {K.~Z.}\ \bibnamefont
  {{H}atsagortsyan}}, \bibinfo {author} {\bibfnamefont {C.~H.}\ \bibnamefont
  {{K}eitel}},\ and\ \bibinfo {author} {\bibfnamefont {J.-X.}\ \bibnamefont
  {{L}i}},\ }\bibfield  {title} {\enquote {\bibinfo {title}
  {{U}ltrarelativistic polarized positron jets via collision of electron and
  ultraintense laser beams},}\ }\href
  {https://doi.org/10.1016/j.physletb.2019.135120} {\bibfield  {journal}
  {\bibinfo  {journal} {{P}hysics {L}etters {B}}\ }\textbf {\bibinfo {volume}
  {800}},\ \bibinfo {pages} {135120} (\bibinfo {year} {2020})}\BibitemShut
  {NoStop}%
\bibitem [{\citenamefont {{G}ong}, \citenamefont {{H}atsagortsyan},\ and\
  \citenamefont {{K}eitel}(2021)}]{gong2021spin}%
  \BibitemOpen
  \bibfield  {author} {\bibinfo {author} {\bibfnamefont {Z.}~\bibnamefont
  {{G}ong}}, \bibinfo {author} {\bibfnamefont {K.~Z.}\ \bibnamefont
  {{H}atsagortsyan}},\ and\ \bibinfo {author} {\bibfnamefont {C.~H.}\
  \bibnamefont {{K}eitel}},\ }\bibfield  {title} {\enquote {\bibinfo {title}
  {{R}etrieving {T}ransient {M}agnetic {F}ields of {U}ltrarelativistic {L}aser
  {P}lasma via {E}jected {E}lectron {P}olarization},}\ }\href
  {https://doi.org/10.1103/physrevlett.127.165002} {\bibfield  {journal}
  {\bibinfo  {journal} {{P}hysical {R}eview {L}etters}\ }\textbf {\bibinfo
  {volume} {127}},\ \bibinfo {pages} {165002} (\bibinfo {year}
  {2021})}\BibitemShut {NoStop}%
\bibitem [{\citenamefont {Nerush}\ and\ \citenamefont
  {Kostyukov}(2007)}]{nerush2007radiation}%
  \BibitemOpen
  \bibfield  {author} {\bibinfo {author} {\bibfnamefont {E.~N.}\ \bibnamefont
  {Nerush}}\ and\ \bibinfo {author} {\bibfnamefont {I.~{\relax Yu}.}\
  \bibnamefont {Kostyukov}},\ }\bibfield  {title} {\enquote {\bibinfo {title}
  {Radiation emission by extreme relativistic electrons and pair production by
  hard photons in a strong plasma wakefield},}\ }\href
  {https://doi.org/10.1103/PhysRevE.75.057401} {\bibfield  {journal} {\bibinfo
  {journal} {Physical Review E}\ }\textbf {\bibinfo {volume} {75}},\ \bibinfo
  {pages} {057401} (\bibinfo {year} {2007})}\BibitemShut {NoStop}%
\bibitem [{\citenamefont {Bell}\ and\ \citenamefont {Kirk}(2008)}]{Bell2008}%
  \BibitemOpen
  \bibfield  {author} {\bibinfo {author} {\bibfnamefont {A.~R.}\ \bibnamefont
  {Bell}}\ and\ \bibinfo {author} {\bibfnamefont {J.~G.}\ \bibnamefont
  {Kirk}},\ }\bibfield  {title} {\enquote {\bibinfo {title} {Possibility of
  prolific pair production with high-power lasers},}\ }\href
  {https://doi.org/10.1103/PhysRevLett.101.200403} {\bibfield  {journal}
  {\bibinfo  {journal} {Physical Review Letters}\ }\textbf {\bibinfo {volume}
  {101}},\ \bibinfo {pages} {200403} (\bibinfo {year} {2008})}\BibitemShut
  {NoStop}%
\bibitem [{\citenamefont {Nerush}\ \emph {et~al.}(2011)\citenamefont {Nerush},
  \citenamefont {Kostyukov}, \citenamefont {Fedotov}, \citenamefont {Narozhny},
  \citenamefont {Elkina},\ and\ \citenamefont {Ruhl}}]{Nerush2011}%
  \BibitemOpen
  \bibfield  {author} {\bibinfo {author} {\bibfnamefont {E.~N.}\ \bibnamefont
  {Nerush}}, \bibinfo {author} {\bibfnamefont {I.~{\relax Yu}.}\ \bibnamefont
  {Kostyukov}}, \bibinfo {author} {\bibfnamefont {A.~M.}\ \bibnamefont
  {Fedotov}}, \bibinfo {author} {\bibfnamefont {N.~B.}\ \bibnamefont
  {Narozhny}}, \bibinfo {author} {\bibfnamefont {N.~V.}\ \bibnamefont
  {Elkina}},\ and\ \bibinfo {author} {\bibfnamefont {H.}~\bibnamefont {Ruhl}},\
  }\bibfield  {title} {\enquote {\bibinfo {title} {Laser field absorption in
  self-generated electron-positron pair plasma},}\ }\href
  {https://doi.org/10.1103/PhysRevLett.106.035001} {\bibfield  {journal}
  {\bibinfo  {journal} {Physical Review Letters}\ }\textbf {\bibinfo {volume}
  {106}},\ \bibinfo {pages} {035001} (\bibinfo {year} {2011})}\BibitemShut
  {NoStop}%
\bibitem [{\citenamefont {Ridgers}\ \emph {et~al.}(2012)\citenamefont
  {Ridgers}, \citenamefont {Brady}, \citenamefont {Duclous}, \citenamefont
  {Kirk}, \citenamefont {Bennett}, \citenamefont {Arber}, \citenamefont
  {Robinson},\ and\ \citenamefont {Bell}}]{Ridgers2012}%
  \BibitemOpen
  \bibfield  {author} {\bibinfo {author} {\bibfnamefont {C.~P.}\ \bibnamefont
  {Ridgers}}, \bibinfo {author} {\bibfnamefont {C.~S.}\ \bibnamefont {Brady}},
  \bibinfo {author} {\bibfnamefont {R.}~\bibnamefont {Duclous}}, \bibinfo
  {author} {\bibfnamefont {J.~G.}\ \bibnamefont {Kirk}}, \bibinfo {author}
  {\bibfnamefont {K.}~\bibnamefont {Bennett}}, \bibinfo {author} {\bibfnamefont
  {T.~D.}\ \bibnamefont {Arber}}, \bibinfo {author} {\bibfnamefont {A.~P.~L.}\
  \bibnamefont {Robinson}},\ and\ \bibinfo {author} {\bibfnamefont {A.~R.}\
  \bibnamefont {Bell}},\ }\bibfield  {title} {\enquote {\bibinfo {title} {Dense
  electron-positron plasmas and ultraintense $\ensuremath{\gamma}$ rays from
  laser-irradiated solids},}\ }\href
  {https://doi.org/10.1103/PhysRevLett.108.165006} {\bibfield  {journal}
  {\bibinfo  {journal} {Physical Review Letters}\ }\textbf {\bibinfo {volume}
  {108}},\ \bibinfo {pages} {165006} (\bibinfo {year} {2012})}\BibitemShut
  {NoStop}%
\bibitem [{\citenamefont {Narozhny}\ and\ \citenamefont
  {Fedotov}(2015)}]{narozhny2015quantum}%
  \BibitemOpen
  \bibfield  {author} {\bibinfo {author} {\bibfnamefont {N.~B.}\ \bibnamefont
  {Narozhny}}\ and\ \bibinfo {author} {\bibfnamefont {A.~M.}\ \bibnamefont
  {Fedotov}},\ }\bibfield  {title} {\enquote {\bibinfo {title}
  {Quantum-electrodynamic cascades in intense laser fields},}\ }\href
  {https://doi.org/10.3367/ufne.0185.201501i.0103} {\bibfield  {journal}
  {\bibinfo  {journal} {Physics Uspekhi}\ }\textbf {\bibinfo {volume} {58}},\
  \bibinfo {pages} {95} (\bibinfo {year} {2015})}\BibitemShut {NoStop}%
\bibitem [{\citenamefont {Kostyukov}\ and\ \citenamefont
  {Nerush}(2016{\natexlab{a}})}]{Kostyukov2016}%
  \BibitemOpen
  \bibfield  {author} {\bibinfo {author} {\bibfnamefont {I.~{\relax Yu}.}\
  \bibnamefont {Kostyukov}}\ and\ \bibinfo {author} {\bibfnamefont {E.~N.}\
  \bibnamefont {Nerush}},\ }\bibfield  {title} {\enquote {\bibinfo {title}
  {Production and dynamics of positrons in ultrahigh intensity laser-foil
  interactions},}\ }\href {https://doi.org/10.1063/1.4962567} {\bibfield
  {journal} {\bibinfo  {journal} {Physics of Plasmas}\ }\textbf {\bibinfo
  {volume} {23}},\ \bibinfo {pages} {093119} (\bibinfo {year}
  {2016}{\natexlab{a}})}\BibitemShut {NoStop}%
\bibitem [{\citenamefont {Grismayer}\ \emph {et~al.}(2017)\citenamefont
  {Grismayer}, \citenamefont {Vranic}, \citenamefont {Martins}, \citenamefont
  {Fonseca},\ and\ \citenamefont {Silva}}]{grismayer2017seeded}%
  \BibitemOpen
  \bibfield  {author} {\bibinfo {author} {\bibfnamefont {T.}~\bibnamefont
  {Grismayer}}, \bibinfo {author} {\bibfnamefont {M.}~\bibnamefont {Vranic}},
  \bibinfo {author} {\bibfnamefont {J.~L.}\ \bibnamefont {Martins}}, \bibinfo
  {author} {\bibfnamefont {R.~A.}\ \bibnamefont {Fonseca}},\ and\ \bibinfo
  {author} {\bibfnamefont {L.~O.}\ \bibnamefont {Silva}},\ }\bibfield  {title}
  {\enquote {\bibinfo {title} {{Seeded {QED} cascades in counterpropagating
  laser pulses}},}\ }\href {https://doi.org/10.1103/PhysRevE.95.023210}
  {\bibfield  {journal} {\bibinfo  {journal} {Physical Review E}\ }\textbf
  {\bibinfo {volume} {95}},\ \bibinfo {pages} {023210} (\bibinfo {year}
  {2017})}\BibitemShut {NoStop}%
\bibitem [{\citenamefont {Jirka}\ \emph {et~al.}(2017)\citenamefont {Jirka},
  \citenamefont {Klimo}, \citenamefont {Vranic}, \citenamefont {Weber},\ and\
  \citenamefont {Korn}}]{jirka2017qed}%
  \BibitemOpen
  \bibfield  {author} {\bibinfo {author} {\bibfnamefont {M.}~\bibnamefont
  {Jirka}}, \bibinfo {author} {\bibfnamefont {O.}~\bibnamefont {Klimo}},
  \bibinfo {author} {\bibfnamefont {M.}~\bibnamefont {Vranic}}, \bibinfo
  {author} {\bibfnamefont {S.}~\bibnamefont {Weber}},\ and\ \bibinfo {author}
  {\bibfnamefont {G.}~\bibnamefont {Korn}},\ }\bibfield  {title} {\enquote
  {\bibinfo {title} {Qed cascade with 10 pw-class lasers},}\ }\href
  {https://doi.org/10.1038/s41598-017-15747-1} {\bibfield  {journal} {\bibinfo
  {journal} {Scientific Reports}\ }\textbf {\bibinfo {volume} {7}},\ \bibinfo
  {pages} {15302} (\bibinfo {year} {2017})}\BibitemShut {NoStop}%
\bibitem [{\citenamefont {Luo}\ \emph {et~al.}(2018{\natexlab{a}})\citenamefont
  {Luo}, \citenamefont {Liu}, \citenamefont {Yuan}, \citenamefont {Chen},
  \citenamefont {Yu}, \citenamefont {Li}, \citenamefont {Del~Sorbo},
  \citenamefont {Ridgers},\ and\ \citenamefont {Sheng}}]{luo2018qed}%
  \BibitemOpen
  \bibfield  {author} {\bibinfo {author} {\bibfnamefont {W.}~\bibnamefont
  {Luo}}, \bibinfo {author} {\bibfnamefont {W.-Y.}\ \bibnamefont {Liu}},
  \bibinfo {author} {\bibfnamefont {T.}~\bibnamefont {Yuan}}, \bibinfo {author}
  {\bibfnamefont {M.}~\bibnamefont {Chen}}, \bibinfo {author} {\bibfnamefont
  {J.-Y.}\ \bibnamefont {Yu}}, \bibinfo {author} {\bibfnamefont {F.-Y.}\
  \bibnamefont {Li}}, \bibinfo {author} {\bibfnamefont {D.}~\bibnamefont
  {Del~Sorbo}}, \bibinfo {author} {\bibfnamefont {C.}~\bibnamefont {Ridgers}},\
  and\ \bibinfo {author} {\bibfnamefont {Z.-M.}\ \bibnamefont {Sheng}},\
  }\bibfield  {title} {\enquote {\bibinfo {title} {Qed cascade saturation in
  extreme high fields},}\ }\href {https://doi.org/10.1038/s41598-018-26785-8}
  {\bibfield  {journal} {\bibinfo  {journal} {Scientific Reports}\ }\textbf
  {\bibinfo {volume} {8}},\ \bibinfo {pages} {8400} (\bibinfo {year}
  {2018}{\natexlab{a}})}\BibitemShut {NoStop}%
\bibitem [{\citenamefont {Yuan}\ \emph {et~al.}(2018)\citenamefont {Yuan},
  \citenamefont {Yu}, \citenamefont {Liu}, \citenamefont {Weng}, \citenamefont
  {Yuan}, \citenamefont {Luo}, \citenamefont {Chen}, \citenamefont {Sheng},\
  and\ \citenamefont {Zhang}}]{Yuan2018}%
  \BibitemOpen
  \bibfield  {author} {\bibinfo {author} {\bibfnamefont {T.}~\bibnamefont
  {Yuan}}, \bibinfo {author} {\bibfnamefont {J.~Y.}\ \bibnamefont {Yu}},
  \bibinfo {author} {\bibfnamefont {W.~Y.}\ \bibnamefont {Liu}}, \bibinfo
  {author} {\bibfnamefont {S.~M.}\ \bibnamefont {Weng}}, \bibinfo {author}
  {\bibfnamefont {X.~H.}\ \bibnamefont {Yuan}}, \bibinfo {author}
  {\bibfnamefont {W.}~\bibnamefont {Luo}}, \bibinfo {author} {\bibfnamefont
  {M.}~\bibnamefont {Chen}}, \bibinfo {author} {\bibfnamefont {Z.~M.}\
  \bibnamefont {Sheng}},\ and\ \bibinfo {author} {\bibfnamefont
  {J.}~\bibnamefont {Zhang}},\ }\bibfield  {title} {\enquote {\bibinfo {title}
  {Spatiotemporal distributions of pair production and cascade in solid targets
  irradiated by ultra-relativistic lasers with different polarizations},}\
  }\href {https://doi.org/10.1088/1361-6587/aab3ba} {\bibfield  {journal}
  {\bibinfo  {journal} {Plasma Physics and Controlled Fusion}\ }\textbf
  {\bibinfo {volume} {60}},\ \bibinfo {pages} {065003} (\bibinfo {year}
  {2018})}\BibitemShut {NoStop}%
\bibitem [{\citenamefont {Del~Sorbo}\ \emph {et~al.}(2018)\citenamefont
  {Del~Sorbo}, \citenamefont {Blackman}, \citenamefont {Capdessus},
  \citenamefont {Small}, \citenamefont {Slade-Lowther}, \citenamefont {Luo},
  \citenamefont {Duff}, \citenamefont {Robinson}, \citenamefont {McKenna},
  \citenamefont {Sheng} \emph {et~al.}}]{del2018ion}%
  \BibitemOpen
  \bibfield  {author} {\bibinfo {author} {\bibfnamefont {D.}~\bibnamefont
  {Del~Sorbo}}, \bibinfo {author} {\bibfnamefont {D.~R.}\ \bibnamefont
  {Blackman}}, \bibinfo {author} {\bibfnamefont {R.}~\bibnamefont {Capdessus}},
  \bibinfo {author} {\bibfnamefont {K.}~\bibnamefont {Small}}, \bibinfo
  {author} {\bibfnamefont {C.}~\bibnamefont {Slade-Lowther}}, \bibinfo {author}
  {\bibfnamefont {W.}~\bibnamefont {Luo}}, \bibinfo {author} {\bibfnamefont
  {M.~J.}\ \bibnamefont {Duff}}, \bibinfo {author} {\bibfnamefont {A.~P.~L.}\
  \bibnamefont {Robinson}}, \bibinfo {author} {\bibfnamefont {P.}~\bibnamefont
  {McKenna}}, \bibinfo {author} {\bibfnamefont {Z.~M.}\ \bibnamefont {Sheng}},
  \emph {et~al.},\ }\bibfield  {title} {\enquote {\bibinfo {title} {Efficient
  ion acceleration and dense electron--positron plasma creation in ultra-high
  intensity laser-solid interactions},}\ }\href
  {https://doi.org/10.1088/1367-2630/aaae61} {\bibfield  {journal} {\bibinfo
  {journal} {New Journal of Physics}\ }\textbf {\bibinfo {volume} {20}},\
  \bibinfo {pages} {033014} (\bibinfo {year} {2018})}\BibitemShut {NoStop}%
\bibitem [{\citenamefont {Lu}\ \emph {et~al.}(2018)\citenamefont {Lu},
  \citenamefont {Yu}, \citenamefont {Hu}, \citenamefont {Ge}, \citenamefont
  {Wang}, \citenamefont {Liu}, \citenamefont {Liu}, \citenamefont {Yin},\ and\
  \citenamefont {Shao}}]{Lu2018}%
  \BibitemOpen
  \bibfield  {author} {\bibinfo {author} {\bibfnamefont {Y.}~\bibnamefont
  {Lu}}, \bibinfo {author} {\bibfnamefont {T.-P.}\ \bibnamefont {Yu}}, \bibinfo
  {author} {\bibfnamefont {L.-X.}\ \bibnamefont {Hu}}, \bibinfo {author}
  {\bibfnamefont {Z.-Y.}\ \bibnamefont {Ge}}, \bibinfo {author} {\bibfnamefont
  {W.-Q.}\ \bibnamefont {Wang}}, \bibinfo {author} {\bibfnamefont {J.-X.}\
  \bibnamefont {Liu}}, \bibinfo {author} {\bibfnamefont {K.}~\bibnamefont
  {Liu}}, \bibinfo {author} {\bibfnamefont {Y.}~\bibnamefont {Yin}},\ and\
  \bibinfo {author} {\bibfnamefont {F.-Q.}\ \bibnamefont {Shao}},\ }\bibfield
  {title} {\enquote {\bibinfo {title} {Enhanced copious electron–positron
  pair production via electron injection from a mass-limited foil},}\ }\href
  {https://doi.org/10.1088/1361-6587/aae819} {\bibfield  {journal} {\bibinfo
  {journal} {Plasma Physics and Controlled Fusion}\ }\textbf {\bibinfo {volume}
  {60}},\ \bibinfo {pages} {125008} (\bibinfo {year} {2018})}\BibitemShut
  {NoStop}%
\bibitem [{\citenamefont {Luo}\ \emph {et~al.}(2018{\natexlab{b}})\citenamefont
  {Luo}, \citenamefont {Wu}, \citenamefont {Liu}, \citenamefont {Ma},
  \citenamefont {Li}, \citenamefont {Yuan}, \citenamefont {Yu}, \citenamefont
  {Chen},\ and\ \citenamefont {Sheng}}]{Luo2018}%
  \BibitemOpen
  \bibfield  {author} {\bibinfo {author} {\bibfnamefont {W.}~\bibnamefont
  {Luo}}, \bibinfo {author} {\bibfnamefont {S.-D.}\ \bibnamefont {Wu}},
  \bibinfo {author} {\bibfnamefont {W.-Y.}\ \bibnamefont {Liu}}, \bibinfo
  {author} {\bibfnamefont {Y.-Y.}\ \bibnamefont {Ma}}, \bibinfo {author}
  {\bibfnamefont {F.-Y.}\ \bibnamefont {Li}}, \bibinfo {author} {\bibfnamefont
  {T.}~\bibnamefont {Yuan}}, \bibinfo {author} {\bibfnamefont {J.-Y.}\
  \bibnamefont {Yu}}, \bibinfo {author} {\bibfnamefont {M.}~\bibnamefont
  {Chen}},\ and\ \bibinfo {author} {\bibfnamefont {Z.-M.}\ \bibnamefont
  {Sheng}},\ }\bibfield  {title} {\enquote {\bibinfo {title} {Enhanced
  electron-positron pair production by two obliquely incident lasers
  interacting with a solid target},}\ }\href
  {https://doi.org/10.1088/1361-6587/aad211} {\bibfield  {journal} {\bibinfo
  {journal} {Plasma Physics and Controlled Fusion}\ }\textbf {\bibinfo {volume}
  {60}},\ \bibinfo {pages} {095006} (\bibinfo {year}
  {2018}{\natexlab{b}})}\BibitemShut {NoStop}%
\bibitem [{\citenamefont {Efimenko}\ \emph {et~al.}(2019)\citenamefont
  {Efimenko}, \citenamefont {Bashinov}, \citenamefont {Gonoskov}, \citenamefont
  {Bastrakov}, \citenamefont {Muraviev}, \citenamefont {Meyerov}, \citenamefont
  {Kim},\ and\ \citenamefont {Sergeev}}]{efimenko2019laser}%
  \BibitemOpen
  \bibfield  {author} {\bibinfo {author} {\bibfnamefont {E.~S.}\ \bibnamefont
  {Efimenko}}, \bibinfo {author} {\bibfnamefont {A.~V.}\ \bibnamefont
  {Bashinov}}, \bibinfo {author} {\bibfnamefont {A.~A.}\ \bibnamefont
  {Gonoskov}}, \bibinfo {author} {\bibfnamefont {S.~I.}\ \bibnamefont
  {Bastrakov}}, \bibinfo {author} {\bibfnamefont {A.~A.}\ \bibnamefont
  {Muraviev}}, \bibinfo {author} {\bibfnamefont {I.~B.}\ \bibnamefont
  {Meyerov}}, \bibinfo {author} {\bibfnamefont {A.~V.}\ \bibnamefont {Kim}},\
  and\ \bibinfo {author} {\bibfnamefont {A.~M.}\ \bibnamefont {Sergeev}},\
  }\bibfield  {title} {\enquote {\bibinfo {title} {Laser-driven plasma pinching
  in ${e}^{\ensuremath{-}}{e}^{+}$ cascade},}\ }\href
  {https://doi.org/10.1103/PhysRevE.99.031201} {\bibfield  {journal} {\bibinfo
  {journal} {Physical Review E}\ }\textbf {\bibinfo {volume} {99}},\ \bibinfo
  {pages} {031201} (\bibinfo {year} {2019})}\BibitemShut {NoStop}%
\bibitem [{\citenamefont {Samsonov}, \citenamefont {Nerush},\ and\
  \citenamefont {Kostyukov}(2019)}]{samsonov2019laser}%
  \BibitemOpen
  \bibfield  {author} {\bibinfo {author} {\bibfnamefont {A.~S.}\ \bibnamefont
  {Samsonov}}, \bibinfo {author} {\bibfnamefont {E.~N.}\ \bibnamefont
  {Nerush}},\ and\ \bibinfo {author} {\bibfnamefont {I.~{\relax Yu}.}\
  \bibnamefont {Kostyukov}},\ }\bibfield  {title} {\enquote {\bibinfo {title}
  {Laser-driven vacuum breakdown waves},}\ }\href
  {https://doi.org/10.1038/s41598-019-47355-6} {\bibfield  {journal} {\bibinfo
  {journal} {Scientific Reports}\ }\textbf {\bibinfo {volume} {9}},\ \bibinfo
  {pages} {11133} (\bibinfo {year} {2019})}\BibitemShut {NoStop}%
\bibitem [{\citenamefont {Samsonov}, \citenamefont {Kostyukov},\ and\
  \citenamefont {Nerush}(2021)}]{samsonov2021hydrodynamical}%
  \BibitemOpen
  \bibfield  {author} {\bibinfo {author} {\bibfnamefont {A.~S.}\ \bibnamefont
  {Samsonov}}, \bibinfo {author} {\bibfnamefont {I.~{\relax Yu}.}\ \bibnamefont
  {Kostyukov}},\ and\ \bibinfo {author} {\bibfnamefont {E.~N.}\ \bibnamefont
  {Nerush}},\ }\bibfield  {title} {\enquote {\bibinfo {title} {Hydrodynamical
  model of {QED} cascade expansion in an extremely strong laser pulse},}\
  }\href {https://doi.org/10.1063/5.0035347} {\bibfield  {journal} {\bibinfo
  {journal} {Matter and Radiation at Extremes}\ }\textbf {\bibinfo {volume}
  {6}},\ \bibinfo {pages} {034401} (\bibinfo {year} {2021})}\BibitemShut
  {NoStop}%
\bibitem [{FAC(2016)}]{FACET}%
  \BibitemOpen
  \href {https://doi.org/10.2172/1340171} {\emph {\bibinfo {title} {{Technical
  Design Report for the FACET-II Project at SLAC National Accelerator
  Laboratory}}}} (\bibinfo {year} {2016})\BibitemShut {NoStop}%
\bibitem [{\citenamefont {Nikishov}\ and\ \citenamefont
  {Ritus}(1964)}]{nikishov1964quantum}%
  \BibitemOpen
  \bibfield  {author} {\bibinfo {author} {\bibfnamefont {A.}~\bibnamefont
  {Nikishov}}\ and\ \bibinfo {author} {\bibfnamefont {V.}~\bibnamefont
  {Ritus}},\ }\bibfield  {title} {\enquote {\bibinfo {title} {Quantum processes
  in the field of a plane electromagnetic wave and in a constant field i},}\
  }\href@noop {} {\bibfield  {journal} {\bibinfo  {journal} {Sov. Phys. JETP}\
  }\textbf {\bibinfo {volume} {19}},\ \bibinfo {pages} {529--541} (\bibinfo
  {year} {1964})}\BibitemShut {NoStop}%
\bibitem [{\citenamefont {Berestetskii}, \citenamefont {Lifshitz},\ and\
  \citenamefont {Pitaevskii}(1982)}]{berestetskii1982quantum}%
  \BibitemOpen
  \bibfield  {author} {\bibinfo {author} {\bibfnamefont {V.~B.}\ \bibnamefont
  {Berestetskii}}, \bibinfo {author} {\bibfnamefont {E.~M.}\ \bibnamefont
  {Lifshitz}},\ and\ \bibinfo {author} {\bibfnamefont {L.~P.}\ \bibnamefont
  {Pitaevskii}},\ }\href {https://doi.org/10.1016/C2009-0-24486-2} {\emph
  {\bibinfo {title} {Quantum electrodynamics}}}\ (\bibinfo  {publisher}
  {Butterworth-Heinemann},\ \bibinfo {year} {1982})\BibitemShut {NoStop}%
\bibitem [{\citenamefont {Ritus}(1985)}]{ritus1985quantum}%
  \BibitemOpen
  \bibfield  {author} {\bibinfo {author} {\bibfnamefont {V.}~\bibnamefont
  {Ritus}},\ }\bibfield  {title} {\enquote {\bibinfo {title} {Quantum effects
  of the interaction of elementary particles with an intense electromagnetic
  field},}\ }\href {https://doi.org/10.1007/BF01120220} {\bibfield  {journal}
  {\bibinfo  {journal} {J. Sov. Laser Res.;(United States)}\ }\textbf {\bibinfo
  {volume} {6}} (\bibinfo {year} {1985}),\ 10.1007/BF01120220}\BibitemShut
  {NoStop}%
\bibitem [{\citenamefont {Khokonov}\ and\ \citenamefont
  {Nitta}(2002)}]{khokonov2002standard}%
  \BibitemOpen
  \bibfield  {author} {\bibinfo {author} {\bibfnamefont {M.~K.}\ \bibnamefont
  {Khokonov}}\ and\ \bibinfo {author} {\bibfnamefont {H.}~\bibnamefont
  {Nitta}},\ }\bibfield  {title} {\enquote {\bibinfo {title} {Standard
  radiation spectrum of relativistic electrons: Beyond the synchrotron
  approximation},}\ }\href {https://doi.org/10.1103/PhysRevLett.89.094801}
  {\bibfield  {journal} {\bibinfo  {journal} {Physical review letters}\
  }\textbf {\bibinfo {volume} {89}},\ \bibinfo {pages} {094801} (\bibinfo
  {year} {2002})}\BibitemShut {NoStop}%
\bibitem [{\citenamefont {Ilderton}, \citenamefont {King},\ and\ \citenamefont
  {Seipt}(2019)}]{ilderton2019extended}%
  \BibitemOpen
  \bibfield  {author} {\bibinfo {author} {\bibfnamefont {A.}~\bibnamefont
  {Ilderton}}, \bibinfo {author} {\bibfnamefont {B.}~\bibnamefont {King}},\
  and\ \bibinfo {author} {\bibfnamefont {D.}~\bibnamefont {Seipt}},\ }\bibfield
   {title} {\enquote {\bibinfo {title} {Extended locally constant field
  approximation for nonlinear compton scattering},}\ }\href
  {https://doi.org/10.1103/PhysRevA.99.042121} {\bibfield  {journal} {\bibinfo
  {journal} {Physical Review A}\ }\textbf {\bibinfo {volume} {99}},\ \bibinfo
  {pages} {042121} (\bibinfo {year} {2019})}\BibitemShut {NoStop}%
\bibitem [{\citenamefont {Heinzl}, \citenamefont {King},\ and\ \citenamefont
  {MacLeod}(2020)}]{heinzl2020locally}%
  \BibitemOpen
  \bibfield  {author} {\bibinfo {author} {\bibfnamefont {T.}~\bibnamefont
  {Heinzl}}, \bibinfo {author} {\bibfnamefont {B.}~\bibnamefont {King}},\ and\
  \bibinfo {author} {\bibfnamefont {A.}~\bibnamefont {MacLeod}},\ }\bibfield
  {title} {\enquote {\bibinfo {title} {Locally monochromatic approximation to
  qed in intense laser fields},}\ }\href
  {https://doi.org/10.1103/PhysRevA.102.063110} {\bibfield  {journal} {\bibinfo
   {journal} {Physical Review A}\ }\textbf {\bibinfo {volume} {102}},\ \bibinfo
  {pages} {063110} (\bibinfo {year} {2020})}\BibitemShut {NoStop}%
\bibitem [{\citenamefont {Gelfer}\ \emph {et~al.}(2022)\citenamefont {Gelfer},
  \citenamefont {Fedotov}, \citenamefont {Mironov},\ and\ \citenamefont
  {Weber}}]{gelfer2022nonlinear}%
  \BibitemOpen
  \bibfield  {author} {\bibinfo {author} {\bibfnamefont {E.}~\bibnamefont
  {Gelfer}}, \bibinfo {author} {\bibfnamefont {A.}~\bibnamefont {Fedotov}},
  \bibinfo {author} {\bibfnamefont {A.}~\bibnamefont {Mironov}},\ and\ \bibinfo
  {author} {\bibfnamefont {S.}~\bibnamefont {Weber}},\ }\bibfield  {title}
  {\enquote {\bibinfo {title} {Nonlinear compton scattering in time-dependent
  electric fields: Lcfa and beyond},}\ }\href@noop {} {\bibfield  {journal}
  {\bibinfo  {journal} {arXiv preprint arXiv:2206.08211}\ } (\bibinfo {year}
  {2022})}\BibitemShut {NoStop}%
\bibitem [{\citenamefont {Podszus}\ and\ \citenamefont
  {Di~Piazza}(2019)}]{Podszus19}%
  \BibitemOpen
  \bibfield  {author} {\bibinfo {author} {\bibfnamefont {T.}~\bibnamefont
  {Podszus}}\ and\ \bibinfo {author} {\bibfnamefont {A.}~\bibnamefont
  {Di~Piazza}},\ }\bibfield  {title} {\enquote {\bibinfo {title} {High-energy
  behavior of strong-field qed in an intense plane wave},}\ }\href@noop {}
  {\bibfield  {journal} {\bibinfo  {journal} {Physical Review D}\ }\textbf
  {\bibinfo {volume} {99}},\ \bibinfo {pages} {076004} (\bibinfo {year}
  {2019})}\BibitemShut {NoStop}%
\bibitem [{\citenamefont {Artemenko}, \citenamefont {Nerush},\ and\
  \citenamefont {Kostyukov}(2020)}]{Artemenko20}%
  \BibitemOpen
  \bibfield  {author} {\bibinfo {author} {\bibfnamefont {I.~I.}\ \bibnamefont
  {Artemenko}}, \bibinfo {author} {\bibfnamefont {E.~N.}\ \bibnamefont
  {Nerush}},\ and\ \bibinfo {author} {\bibfnamefont {I.}~\bibnamefont
  {Kostyukov}},\ }\bibfield  {title} {\enquote {\bibinfo {title}
  {Quasiclassical approach to synergic synchrotron-cherenkov radiation in
  polarized vacuum},}\ }\href {https://doi.org/10.1088/1367-2630/abb388}
  {\bibfield  {journal} {\bibinfo  {journal} {New Journal of Physics}\ }\textbf
  {\bibinfo {volume} {22}},\ \bibinfo {pages} {093072} (\bibinfo {year}
  {2020})}\BibitemShut {NoStop}%
\bibitem [{\citenamefont {Kirk}, \citenamefont {Bell},\ and\ \citenamefont
  {Arka}(2009)}]{kirk2009pair}%
  \BibitemOpen
  \bibfield  {author} {\bibinfo {author} {\bibfnamefont {J.~G.}\ \bibnamefont
  {Kirk}}, \bibinfo {author} {\bibfnamefont {A.}~\bibnamefont {Bell}},\ and\
  \bibinfo {author} {\bibfnamefont {I.}~\bibnamefont {Arka}},\ }\bibfield
  {title} {\enquote {\bibinfo {title} {Pair production in counter-propagating
  laser beams},}\ }\href {https://doi.org/10.1088/0741-3335/51/8/085008}
  {\bibfield  {journal} {\bibinfo  {journal} {Plasma Physics and Controlled
  Fusion}\ }\textbf {\bibinfo {volume} {51}},\ \bibinfo {pages} {085008}
  (\bibinfo {year} {2009})}\BibitemShut {NoStop}%
\bibitem [{\citenamefont {Bulanov}\ \emph {et~al.}(2013)\citenamefont
  {Bulanov}, \citenamefont {Schroeder}, \citenamefont {Esarey},\ and\
  \citenamefont {Leemans}}]{bulanov2013electromagnetic}%
  \BibitemOpen
  \bibfield  {author} {\bibinfo {author} {\bibfnamefont {S.}~\bibnamefont
  {Bulanov}}, \bibinfo {author} {\bibfnamefont {C.}~\bibnamefont {Schroeder}},
  \bibinfo {author} {\bibfnamefont {E.}~\bibnamefont {Esarey}},\ and\ \bibinfo
  {author} {\bibfnamefont {W.}~\bibnamefont {Leemans}},\ }\bibfield  {title}
  {\enquote {\bibinfo {title} {Electromagnetic cascade in high-energy electron,
  positron, and photon interactions with intense laser pulses},}\ }\href
  {https://doi.org/10.1103/PhysRevA.87.062110} {\bibfield  {journal} {\bibinfo
  {journal} {Physical Review A}\ }\textbf {\bibinfo {volume} {87}},\ \bibinfo
  {pages} {062110} (\bibinfo {year} {2013})}\BibitemShut {NoStop}%
\bibitem [{\citenamefont {Esirkepov}\ \emph {et~al.}(2015)\citenamefont
  {Esirkepov}, \citenamefont {Bulanov}, \citenamefont {Koga}, \citenamefont
  {Kando}, \citenamefont {Kondo}, \citenamefont {Rosanov}, \citenamefont
  {Korn},\ and\ \citenamefont {Bulanov}}]{esirkepov2014attractors}%
  \BibitemOpen
  \bibfield  {author} {\bibinfo {author} {\bibfnamefont {T.~Z.}\ \bibnamefont
  {Esirkepov}}, \bibinfo {author} {\bibfnamefont {S.~S.}\ \bibnamefont
  {Bulanov}}, \bibinfo {author} {\bibfnamefont {J.~K.}\ \bibnamefont {Koga}},
  \bibinfo {author} {\bibfnamefont {M.}~\bibnamefont {Kando}}, \bibinfo
  {author} {\bibfnamefont {K.}~\bibnamefont {Kondo}}, \bibinfo {author}
  {\bibfnamefont {N.~N.}\ \bibnamefont {Rosanov}}, \bibinfo {author}
  {\bibfnamefont {G.}~\bibnamefont {Korn}},\ and\ \bibinfo {author}
  {\bibfnamefont {S.~V.}\ \bibnamefont {Bulanov}},\ }\bibfield  {title}
  {\enquote {\bibinfo {title} {Attractors and chaos of electron dynamics in
  electromagnetic standing wave},}\ }\href
  {https://doi.org/10.1016/j.physleta.2015.06.017} {\bibfield  {journal}
  {\bibinfo  {journal} {Physics Letters A}\ }\textbf {\bibinfo {volume}
  {379}},\ \bibinfo {pages} {2044} (\bibinfo {year} {2015})}\BibitemShut
  {NoStop}%
\bibitem [{\citenamefont {Niel}\ \emph {et~al.}(2018)\citenamefont {Niel},
  \citenamefont {Riconda}, \citenamefont {Amiranoff}, \citenamefont {Duclous},\
  and\ \citenamefont {Grech}}]{niel2018quantum}%
  \BibitemOpen
  \bibfield  {author} {\bibinfo {author} {\bibfnamefont {F.}~\bibnamefont
  {Niel}}, \bibinfo {author} {\bibfnamefont {C.}~\bibnamefont {Riconda}},
  \bibinfo {author} {\bibfnamefont {F.}~\bibnamefont {Amiranoff}}, \bibinfo
  {author} {\bibfnamefont {R.}~\bibnamefont {Duclous}},\ and\ \bibinfo {author}
  {\bibfnamefont {M.}~\bibnamefont {Grech}},\ }\bibfield  {title} {\enquote
  {\bibinfo {title} {From quantum to classical modeling of radiation reaction:
  A focus on stochasticity effects},}\ }\href
  {https://doi.org/10.1103/physreve.97.043209} {\bibfield  {journal} {\bibinfo
  {journal} {Physical Review E}\ }\textbf {\bibinfo {volume} {97}},\ \bibinfo
  {pages} {043209} (\bibinfo {year} {2018})}\BibitemShut {NoStop}%
\bibitem [{\citenamefont {Gonoskov}\ \emph {et~al.}(2021)\citenamefont
  {Gonoskov}, \citenamefont {Blackburn}, \citenamefont {Marklund},\ and\
  \citenamefont {Bulanov}}]{gonoskov2021charged}%
  \BibitemOpen
  \bibfield  {author} {\bibinfo {author} {\bibfnamefont {A.}~\bibnamefont
  {Gonoskov}}, \bibinfo {author} {\bibfnamefont {T.}~\bibnamefont {Blackburn}},
  \bibinfo {author} {\bibfnamefont {M.}~\bibnamefont {Marklund}},\ and\
  \bibinfo {author} {\bibfnamefont {S.}~\bibnamefont {Bulanov}},\ }\bibfield
  {title} {\enquote {\bibinfo {title} {Charged particle motion and radiation in
  strong electromagnetic fields},}\ }\href@noop {} {\bibfield  {journal}
  {\bibinfo  {journal} {arXiv preprint arXiv:2107.02161}\ } (\bibinfo {year}
  {2021})}\BibitemShut {NoStop}%
\bibitem [{\citenamefont {Shen}\ and\ \citenamefont
  {White}(1972)}]{shen1972energy}%
  \BibitemOpen
  \bibfield  {author} {\bibinfo {author} {\bibfnamefont {C.}~\bibnamefont
  {Shen}}\ and\ \bibinfo {author} {\bibfnamefont {D.}~\bibnamefont {White}},\
  }\bibfield  {title} {\enquote {\bibinfo {title} {Energy straggling and
  radiation reaction for magnetic bremsstrahlung},}\ }\href
  {https://doi.org/10.1103/PhysRevLett.28.455} {\bibfield  {journal} {\bibinfo
  {journal} {Physical Review Letters}\ }\textbf {\bibinfo {volume} {28}},\
  \bibinfo {pages} {455} (\bibinfo {year} {1972})}\BibitemShut {NoStop}%
\bibitem [{\citenamefont {Duclous}, \citenamefont {Kirk},\ and\ \citenamefont
  {Bell}(2010)}]{duclous2010monte}%
  \BibitemOpen
  \bibfield  {author} {\bibinfo {author} {\bibfnamefont {R.}~\bibnamefont
  {Duclous}}, \bibinfo {author} {\bibfnamefont {J.~G.}\ \bibnamefont {Kirk}},\
  and\ \bibinfo {author} {\bibfnamefont {A.~R.}\ \bibnamefont {Bell}},\
  }\bibfield  {title} {\enquote {\bibinfo {title} {Monte carlo calculations of
  pair production in high-intensity laser--plasma interactions},}\ }\href
  {https://doi.org/10.1088/0741-3335/53/1/015009} {\bibfield  {journal}
  {\bibinfo  {journal} {Plasma Physics and Controlled Fusion}\ }\textbf
  {\bibinfo {volume} {53}},\ \bibinfo {pages} {015009} (\bibinfo {year}
  {2010})}\BibitemShut {NoStop}%
\bibitem [{\citenamefont {Harvey}\ \emph {et~al.}(2017)\citenamefont {Harvey},
  \citenamefont {Gonoskov}, \citenamefont {Ilderton},\ and\ \citenamefont
  {Marklund}}]{harvey2017quantum}%
  \BibitemOpen
  \bibfield  {author} {\bibinfo {author} {\bibfnamefont {C.}~\bibnamefont
  {Harvey}}, \bibinfo {author} {\bibfnamefont {A.}~\bibnamefont {Gonoskov}},
  \bibinfo {author} {\bibfnamefont {A.}~\bibnamefont {Ilderton}},\ and\
  \bibinfo {author} {\bibfnamefont {M.}~\bibnamefont {Marklund}},\ }\bibfield
  {title} {\enquote {\bibinfo {title} {Quantum quenching of radiation losses in
  short laser pulses},}\ }\href
  {https://doi.org/10.1103/PhysRevLett.118.105004} {\bibfield  {journal}
  {\bibinfo  {journal} {Physical Review Letters}\ }\textbf {\bibinfo {volume}
  {118}},\ \bibinfo {pages} {105004} (\bibinfo {year} {2017})}\BibitemShut
  {NoStop}%
\bibitem [{\citenamefont {Neitz}\ and\ \citenamefont
  {Di~Piazza}(2013)}]{neitz2013stochasticity}%
  \BibitemOpen
  \bibfield  {author} {\bibinfo {author} {\bibfnamefont {N.}~\bibnamefont
  {Neitz}}\ and\ \bibinfo {author} {\bibfnamefont {A.}~\bibnamefont
  {Di~Piazza}},\ }\bibfield  {title} {\enquote {\bibinfo {title} {Stochasticity
  effects in quantum radiation reaction},}\ }\href
  {https://doi.org/10.1103/PhysRevLett.111.054802} {\bibfield  {journal}
  {\bibinfo  {journal} {Physical review letters}\ }\textbf {\bibinfo {volume}
  {111}},\ \bibinfo {pages} {054802} (\bibinfo {year} {2013})}\BibitemShut
  {NoStop}%
\bibitem [{\citenamefont {Ridgers}\ \emph {et~al.}(2017)\citenamefont
  {Ridgers}, \citenamefont {Blackburn}, \citenamefont {Del~Sorbo},
  \citenamefont {Bradley}, \citenamefont {Slade-Lowther}, \citenamefont
  {Baird}, \citenamefont {Mangles}, \citenamefont {McKenna}, \citenamefont
  {Marklund}, \citenamefont {Murphy} \emph {et~al.}}]{ridgers2017signatures}%
  \BibitemOpen
  \bibfield  {author} {\bibinfo {author} {\bibfnamefont {C.}~\bibnamefont
  {Ridgers}}, \bibinfo {author} {\bibfnamefont {T.}~\bibnamefont {Blackburn}},
  \bibinfo {author} {\bibfnamefont {D.}~\bibnamefont {Del~Sorbo}}, \bibinfo
  {author} {\bibfnamefont {L.}~\bibnamefont {Bradley}}, \bibinfo {author}
  {\bibfnamefont {C.}~\bibnamefont {Slade-Lowther}}, \bibinfo {author}
  {\bibfnamefont {C.}~\bibnamefont {Baird}}, \bibinfo {author} {\bibfnamefont
  {S.}~\bibnamefont {Mangles}}, \bibinfo {author} {\bibfnamefont
  {P.}~\bibnamefont {McKenna}}, \bibinfo {author} {\bibfnamefont
  {M.}~\bibnamefont {Marklund}}, \bibinfo {author} {\bibfnamefont
  {C.}~\bibnamefont {Murphy}}, \emph {et~al.},\ }\bibfield  {title} {\enquote
  {\bibinfo {title} {Signatures of quantum effects on radiation reaction in
  laser--electron-beam collisions},}\ }\href
  {https://doi.org/10.1017/S0022377817000642} {\bibfield  {journal} {\bibinfo
  {journal} {Journal of Plasma Physics}\ }\textbf {\bibinfo {volume} {83}},\
  \bibinfo {pages} {715830502} (\bibinfo {year} {2017})}\BibitemShut {NoStop}%
\bibitem [{\citenamefont {Gerlach}\ and\ \citenamefont
  {Stern}(1922)}]{gerlach1922experimentelle}%
  \BibitemOpen
  \bibfield  {author} {\bibinfo {author} {\bibfnamefont {W.}~\bibnamefont
  {Gerlach}}\ and\ \bibinfo {author} {\bibfnamefont {O.}~\bibnamefont
  {Stern}},\ }\bibfield  {title} {\enquote {\bibinfo {title} {Der
  experimentelle nachweis der richtungsquantelung im magnetfeld},}\ }\href
  {https://doi.org/10.1007/BF01326983} {\bibfield  {journal} {\bibinfo
  {journal} {Zeitschrift f{\"u}r Physik}\ }\textbf {\bibinfo {volume} {9}},\
  \bibinfo {pages} {349--352} (\bibinfo {year} {1922})}\BibitemShut {NoStop}%
\bibitem [{\citenamefont {Thomas}(1926)}]{thomas1926motion}%
  \BibitemOpen
  \bibfield  {author} {\bibinfo {author} {\bibfnamefont {L.~H.}\ \bibnamefont
  {Thomas}},\ }\bibfield  {title} {\enquote {\bibinfo {title} {The motion of
  the spinning electron},}\ }\href {https://doi.org/10.1038/117514a0}
  {\bibfield  {journal} {\bibinfo  {journal} {Nature}\ }\textbf {\bibinfo
  {volume} {117}},\ \bibinfo {pages} {514--514} (\bibinfo {year}
  {1926})}\BibitemShut {NoStop}%
\bibitem [{\citenamefont {Bargmann}, \citenamefont {Michel},\ and\
  \citenamefont {Telegdi}(1959)}]{bargmann1959precession}%
  \BibitemOpen
  \bibfield  {author} {\bibinfo {author} {\bibfnamefont {V.}~\bibnamefont
  {Bargmann}}, \bibinfo {author} {\bibfnamefont {L.}~\bibnamefont {Michel}},\
  and\ \bibinfo {author} {\bibfnamefont {V.}~\bibnamefont {Telegdi}},\
  }\bibfield  {title} {\enquote {\bibinfo {title} {Precession of the
  polarization of particles moving in a homogeneous electromagnetic field},}\
  }\href {https://doi.org/10.1103/PhysRevLett.2.435} {\bibfield  {journal}
  {\bibinfo  {journal} {Physical Review Letters}\ }\textbf {\bibinfo {volume}
  {2}},\ \bibinfo {pages} {435} (\bibinfo {year} {1959})}\BibitemShut {NoStop}%
\bibitem [{\citenamefont {Mane}, \citenamefont {Shatunov},\ and\ \citenamefont
  {Yokoya}(2005)}]{mane2005spin}%
  \BibitemOpen
  \bibfield  {author} {\bibinfo {author} {\bibfnamefont {S.}~\bibnamefont
  {Mane}}, \bibinfo {author} {\bibfnamefont {Y.~M.}\ \bibnamefont {Shatunov}},\
  and\ \bibinfo {author} {\bibfnamefont {K.}~\bibnamefont {Yokoya}},\
  }\bibfield  {title} {\enquote {\bibinfo {title} {Spin-polarized charged
  particle beams in high-energy accelerators},}\ }\href
  {https://doi.org/10.1088/0034-4885/68/9/R01} {\bibfield  {journal} {\bibinfo
  {journal} {Reports on Progress in Physics}\ }\textbf {\bibinfo {volume}
  {68}},\ \bibinfo {pages} {1997} (\bibinfo {year} {2005})}\BibitemShut
  {NoStop}%
\bibitem [{\citenamefont {Seipt}\ \emph {et~al.}(2021)\citenamefont {Seipt},
  \citenamefont {Ridgers}, \citenamefont {Del~Sorbo},\ and\ \citenamefont
  {Thomas}}]{Seipt2021PolarizedQEDcascad}%
  \BibitemOpen
  \bibfield  {author} {\bibinfo {author} {\bibfnamefont {D.}~\bibnamefont
  {Seipt}}, \bibinfo {author} {\bibfnamefont {C.~P.}\ \bibnamefont {Ridgers}},
  \bibinfo {author} {\bibfnamefont {D.}~\bibnamefont {Del~Sorbo}},\ and\
  \bibinfo {author} {\bibfnamefont {A.~G.~R.}\ \bibnamefont {Thomas}},\
  }\bibfield  {title} {\enquote {\bibinfo {title} {Polarized {QED} cascades},}\
  }\href {https://doi.org/10.1088/1367-2630/abf584} {\bibfield  {journal}
  {\bibinfo  {journal} {New Journal of Physics}\ }\textbf {\bibinfo {volume}
  {23}},\ \bibinfo {pages} {053025} (\bibinfo {year} {2021})}\BibitemShut
  {NoStop}%
\bibitem [{\citenamefont {Wen}, \citenamefont {Tamburini},\ and\ \citenamefont
  {Keitel}(2019)}]{wen2019polarized}%
  \BibitemOpen
  \bibfield  {author} {\bibinfo {author} {\bibfnamefont {M.}~\bibnamefont
  {Wen}}, \bibinfo {author} {\bibfnamefont {M.}~\bibnamefont {Tamburini}},\
  and\ \bibinfo {author} {\bibfnamefont {C.~H.}\ \bibnamefont {Keitel}},\
  }\bibfield  {title} {\enquote {\bibinfo {title} {Polarized
  laser-wakefield-accelerated kiloampere electron beams},}\ }\href
  {https://doi.org/10.1103/physrevlett.122.214801} {\bibfield  {journal}
  {\bibinfo  {journal} {Physical review letters}\ }\textbf {\bibinfo {volume}
  {122}},\ \bibinfo {pages} {214801} (\bibinfo {year} {2019})}\BibitemShut
  {NoStop}%
\bibitem [{\citenamefont {Li}\ \emph {et~al.}(2019)\citenamefont {Li},
  \citenamefont {Shaisultanov}, \citenamefont {Hatsagortsyan}, \citenamefont
  {Wan}, \citenamefont {Keitel},\ and\ \citenamefont
  {Li}}]{li2019ultrarelativistic}%
  \BibitemOpen
  \bibfield  {author} {\bibinfo {author} {\bibfnamefont {Y.-F.}\ \bibnamefont
  {Li}}, \bibinfo {author} {\bibfnamefont {R.}~\bibnamefont {Shaisultanov}},
  \bibinfo {author} {\bibfnamefont {K.~Z.}\ \bibnamefont {Hatsagortsyan}},
  \bibinfo {author} {\bibfnamefont {F.}~\bibnamefont {Wan}}, \bibinfo {author}
  {\bibfnamefont {C.~H.}\ \bibnamefont {Keitel}},\ and\ \bibinfo {author}
  {\bibfnamefont {J.-X.}\ \bibnamefont {Li}},\ }\bibfield  {title} {\enquote
  {\bibinfo {title} {Ultrarelativistic electron-beam polarization in
  single-shot interaction with an ultraintense laser pulse},}\ }\href
  {https://doi.org/10.1103/physrevlett.122.154801} {\bibfield  {journal}
  {\bibinfo  {journal} {Physical review letters}\ }\textbf {\bibinfo {volume}
  {122}},\ \bibinfo {pages} {154801} (\bibinfo {year} {2019})}\BibitemShut
  {NoStop}%
\bibitem [{\citenamefont {Gong}, \citenamefont {Hatsagortsyan},\ and\
  \citenamefont {Keitel}(2021)}]{gong2021retrieving}%
  \BibitemOpen
  \bibfield  {author} {\bibinfo {author} {\bibfnamefont {Z.}~\bibnamefont
  {Gong}}, \bibinfo {author} {\bibfnamefont {K.~Z.}\ \bibnamefont
  {Hatsagortsyan}},\ and\ \bibinfo {author} {\bibfnamefont {C.~H.}\
  \bibnamefont {Keitel}},\ }\bibfield  {title} {\enquote {\bibinfo {title}
  {Retrieving transient magnetic fields of ultrarelativistic laser plasma via
  ejected electron polarization},}\ }\href
  {https://doi.org/10.1103/PhysRevLett.127.165002} {\bibfield  {journal}
  {\bibinfo  {journal} {Physical review letters}\ }\textbf {\bibinfo {volume}
  {127}},\ \bibinfo {pages} {165002} (\bibinfo {year} {2021})}\BibitemShut
  {NoStop}%
\bibitem [{\citenamefont {Samsonov}, \citenamefont {Nerush},\ and\
  \citenamefont {Kostyukov}(2018)}]{samsonov2018}%
  \BibitemOpen
  \bibfield  {author} {\bibinfo {author} {\bibfnamefont {A.~S.}\ \bibnamefont
  {Samsonov}}, \bibinfo {author} {\bibfnamefont {E.~N.}\ \bibnamefont
  {Nerush}},\ and\ \bibinfo {author} {\bibfnamefont {I.~{\relax Yu}.}\
  \bibnamefont {Kostyukov}},\ }\bibfield  {title} {\enquote {\bibinfo {title}
  {Asymptotic electron motion in the strongly-radiation-dominated regime},}\
  }\href {https://doi.org/10.1103/PhysRevA.98.053858} {\bibfield  {journal}
  {\bibinfo  {journal} {Physical Review A}\ }\textbf {\bibinfo {volume} {98}},\
  \bibinfo {pages} {053858} (\bibinfo {year} {2018})}\BibitemShut {NoStop}%
\bibitem [{\citenamefont {Gonoskov}\ and\ \citenamefont
  {Marklund}(2018)}]{gonoskov2018radiation}%
  \BibitemOpen
  \bibfield  {author} {\bibinfo {author} {\bibfnamefont {A.}~\bibnamefont
  {Gonoskov}}\ and\ \bibinfo {author} {\bibfnamefont {M.}~\bibnamefont
  {Marklund}},\ }\bibfield  {title} {\enquote {\bibinfo {title}
  {Radiation-dominated particle and plasma dynamics},}\ }\href
  {https://doi.org/10.1063/1.5047799} {\bibfield  {journal} {\bibinfo
  {journal} {Physics of Plasmas}\ }\textbf {\bibinfo {volume} {25}},\ \bibinfo
  {pages} {093109} (\bibinfo {year} {2018})}\BibitemShut {NoStop}%
\bibitem [{\citenamefont {J{\'e}r{\^o}me}(2022)}]{jerome2022particle}%
  \BibitemOpen
  \bibfield  {author} {\bibinfo {author} {\bibfnamefont {P.}~\bibnamefont
  {J{\'e}r{\^o}me}},\ }\bibfield  {title} {\enquote {\bibinfo {title} {Particle
  acceleration and radiation reaction in a strongly magnetized rotating
  dipole},}\ }\href@noop {} {\bibfield  {journal} {\bibinfo  {journal} {arXiv
  preprint arXiv:2207.00624}\ } (\bibinfo {year} {2022})}\BibitemShut {NoStop}%
\bibitem [{\citenamefont {Zel{'}dovich}(1975)}]{Zeldovich75}%
  \BibitemOpen
  \bibfield  {author} {\bibinfo {author} {\bibfnamefont {Y.~B.}\ \bibnamefont
  {Zel{'}dovich}},\ }\bibfield  {title} {\enquote {\bibinfo {title}
  {Interaction of free electrons with electromagnetic radiation},}\ }\href
  {https://doi.org/10.1070/PU1975v018n02ABEH001947} {\bibfield  {journal}
  {\bibinfo  {journal} {Soviet Physics Uspekhi}\ }\textbf {\bibinfo {volume}
  {18}},\ \bibinfo {pages} {79} (\bibinfo {year} {1975})}\BibitemShut {NoStop}%
\bibitem [{\citenamefont {Kostyukov}\ and\ \citenamefont
  {Nerush}(2016{\natexlab{b}})}]{kostyukov2016production}%
  \BibitemOpen
  \bibfield  {author} {\bibinfo {author} {\bibfnamefont {I.~{\relax Yu}.}\
  \bibnamefont {Kostyukov}}\ and\ \bibinfo {author} {\bibfnamefont {E.~N.}\
  \bibnamefont {Nerush}},\ }\bibfield  {title} {\enquote {\bibinfo {title}
  {Production and dynamics of positrons in ultrahigh intensity laser-foil
  interactions},}\ }\href {https://doi.org/10.1063/1.4962567} {\bibfield
  {journal} {\bibinfo  {journal} {Physics of Plasmas}\ }\textbf {\bibinfo
  {volume} {23}},\ \bibinfo {pages} {093119} (\bibinfo {year}
  {2016}{\natexlab{b}})}\BibitemShut {NoStop}%
\bibitem [{\citenamefont {Ekman}, \citenamefont {Heinzl},\ and\ \citenamefont
  {Ilderton}(2021)}]{ekman2021exact}%
  \BibitemOpen
  \bibfield  {author} {\bibinfo {author} {\bibfnamefont {R.}~\bibnamefont
  {Ekman}}, \bibinfo {author} {\bibfnamefont {T.}~\bibnamefont {Heinzl}},\ and\
  \bibinfo {author} {\bibfnamefont {A.}~\bibnamefont {Ilderton}},\ }\bibfield
  {title} {\enquote {\bibinfo {title} {Exact solutions in radiation reaction
  and the radiation-free direction},}\ }\href
  {https://doi.org/10.1088/1367-2630/abfab2} {\bibfield  {journal} {\bibinfo
  {journal} {New Journal of Physics}\ }\textbf {\bibinfo {volume} {23}},\
  \bibinfo {pages} {055001} (\bibinfo {year} {2021})}\BibitemShut {NoStop}%
\bibitem [{\citenamefont {Landau}(2013)}]{landau2013classical}%
  \BibitemOpen
  \bibfield  {author} {\bibinfo {author} {\bibfnamefont {L.~D.}\ \bibnamefont
  {Landau}},\ }\href@noop {} {\emph {\bibinfo {title} {The classical theory of
  fields}}},\ Vol.~\bibinfo {volume} {2}\ (\bibinfo  {publisher} {Elsevier},\
  \bibinfo {year} {2013})\BibitemShut {NoStop}%
\bibitem [{\citenamefont {Di~Piazza}(2008)}]{di2008exact}%
  \BibitemOpen
  \bibfield  {author} {\bibinfo {author} {\bibfnamefont {A.}~\bibnamefont
  {Di~Piazza}},\ }\bibfield  {title} {\enquote {\bibinfo {title} {Exact
  solution of the landau-lifshitz equation in a plane wave},}\ }\href
  {https://doi.org/10.1007/s11005-008-0228-9} {\bibfield  {journal} {\bibinfo
  {journal} {Letters in Mathematical Physics}\ }\textbf {\bibinfo {volume}
  {83}},\ \bibinfo {pages} {305--313} (\bibinfo {year} {2008})}\BibitemShut
  {NoStop}%
\bibitem [{\citenamefont {Gunn}\ and\ \citenamefont
  {Ostriker}(1971)}]{gunn1971motion}%
  \BibitemOpen
  \bibfield  {author} {\bibinfo {author} {\bibfnamefont {J.~E.}\ \bibnamefont
  {Gunn}}\ and\ \bibinfo {author} {\bibfnamefont {J.~P.}\ \bibnamefont
  {Ostriker}},\ }\bibfield  {title} {\enquote {\bibinfo {title} {On the motion
  and radiation of charged particles in strong electromagnetic waves. i. motion
  in plane and spherical waves},}\ }\href {https://doi.org/10.1086/150919}
  {\bibfield  {journal} {\bibinfo  {journal} {The Astrophysical Journal}\
  }\textbf {\bibinfo {volume} {165}},\ \bibinfo {pages} {523} (\bibinfo {year}
  {1971})}\BibitemShut {NoStop}%
\bibitem [{\citenamefont {Grewing}, \citenamefont {Schr{\"u}fer},\ and\
  \citenamefont {Heintzmann}(1973)}]{grewing1973acceleration}%
  \BibitemOpen
  \bibfield  {author} {\bibinfo {author} {\bibfnamefont {M.}~\bibnamefont
  {Grewing}}, \bibinfo {author} {\bibfnamefont {E.}~\bibnamefont
  {Schr{\"u}fer}},\ and\ \bibinfo {author} {\bibfnamefont {H.}~\bibnamefont
  {Heintzmann}},\ }\bibfield  {title} {\enquote {\bibinfo {title} {Acceleration
  of charged particles and radiation reaction in strong plane and spherical
  waves. ii},}\ }\href {https://doi.org/10.1007/BF01397962} {\bibfield
  {journal} {\bibinfo  {journal} {Zeitschrift f{\"u}r Physik A Hadrons and
  nuclei}\ }\textbf {\bibinfo {volume} {260}},\ \bibinfo {pages} {375--384}
  (\bibinfo {year} {1973})}\BibitemShut {NoStop}%
\bibitem [{\citenamefont {Thielheim}(1993)}]{thielheim1993particle}%
  \BibitemOpen
  \bibfield  {author} {\bibinfo {author} {\bibfnamefont {K.}~\bibnamefont
  {Thielheim}},\ }\bibfield  {title} {\enquote {\bibinfo {title} {Particle
  acceleration in extremely strong electromagnetic wave fields},}\ }in\ \href
  {https://doi.org/10.1109/PAC.1993.308941} {\emph {\bibinfo {booktitle}
  {Proceedings of International Conference on Particle Accelerators}}}\
  (\bibinfo {organization} {IEEE},\ \bibinfo {year} {1993})\ pp.\ \bibinfo
  {pages} {276--278}\BibitemShut {NoStop}%
\end{thebibliography}%

\end{document}